\documentclass[11pt,a4paper]{article}

\usepackage[utf8]{inputenc}
\usepackage[T1]{fontenc}
\usepackage[margin=2.5cm]{geometry}
\usepackage{amsmath,amssymb,amsfonts}
\usepackage{mathtools}
\usepackage{graphicx}
\usepackage{xcolor}
\usepackage{booktabs}
\usepackage{multirow}
\usepackage{tabularx}
\usepackage[ruled,vlined,linesnumbered]{algorithm2e}
\usepackage{float}
\usepackage{tikz}
\usetikzlibrary{shapes.geometric, arrows.meta, positioning, fit, backgrounds,
                 calc, decorations.markings, patterns, shadows, matrix}
\usepackage{pgfplots}
\pgfplotsset{compat=1.18}
\usepackage{hyperref}
\hypersetup{colorlinks=true,linkcolor=blue!70!black,citecolor=green!50!black,urlcolor=blue!70!black}
\usepackage{cleveref}
\usepackage{subcaption}
\usepackage{enumitem}
\usepackage{siunitx}
\usepackage{longtable}
\usepackage{caption}
\usepackage{authblk}

\definecolor{agent1}{RGB}{220,50,47}
\definecolor{agent2}{RGB}{38,139,210}
\definecolor{agent3}{RGB}{133,153,0}
\definecolor{agent4}{RGB}{203,75,22}
\definecolor{agent5}{RGB}{108,113,196}
\definecolor{agent6}{RGB}{42,161,152}
\definecolor{agent7}{RGB}{181,137,0}
\definecolor{agent8}{RGB}{211,54,130}
\definecolor{agent9}{RGB}{0,43,54}
\definecolor{llmblue}{RGB}{30,80,160}
\definecolor{pipegray}{RGB}{240,242,245}
\definecolor{passgreen}{RGB}{39,174,96}
\definecolor{failred}{RGB}{192,57,43}
\definecolor{warnoran}{RGB}{243,156,18}

\newcommand{\rev}[1]{\textcolor{black}{#1}}

\newcommand{\bx}{\mathbf{x}}
\newcommand{\bu}{\mathbf{u}}
\newcommand{\bv}{\mathbf{v}}
\newcommand{\bF}{\mathbf{F}}
\newcommand{\bff}{\mathbf{f}}
\newcommand{\bb}{\mathbf{b}}
\newcommand{\bn}{\mathbf{n}}
\newcommand{\bt}{\mathbf{t}}
\newcommand{\bsig}{\boldsymbol{\sigma}}
\newcommand{\beps}{\boldsymbol{\varepsilon}}
\newcommand{\bC}{\mathbb{C}}
\newcommand{\R}{\mathbb{R}}
\newcommand{\dif}{\,\mathrm{d}}

\begin{document}

\title{From Perception to Autonomous Computational Modeling: A Multi-Agent Approach}

\author[1]{Daniel N.~Wilke\thanks{Corresponding author. Email: \texttt{daniel.wilke@wits.ac.za}}}
\affil[1]{School of Mechanical, Industrial and Aeronautical Engineering, University of the Witwatersrand, Johannesburg, South Africa}

\date{\today}
\maketitle

\begin{abstract}
We present a solver-agnostic framework in which coordinated large language model (LLM) agents autonomously execute the complete computational mechanics workflow, from perceptual data of an engineering component through geometry extraction, material inference, discretisation, solver execution, uncertainty quantification, and code-compliant assessment, to an engineering report with actionable recommendations. Agents are formalised as conditioned operators on a shared context space with quality gates that introduce conditional iteration between pipeline layers. We introduce a mathematical framework for extracting engineering information from perceptual data under uncertainty using interval bounds, probability densities, and fuzzy membership functions, \rev{treating the bands attached to image-extracted parameters as first-iteration self-reported plausibility tags rather than empirically calibrated confidence intervals, because ground truth is not recoverable from a single photograph and must be supplied by the supervising engineer at review time}, and introduce \emph{task-dependent conservatism} to resolve the ambiguity of what ``conservative'' means when different limit states are governed by opposing parameter trends. The framework is demonstrated through a finite element analysis pipeline applied to a photograph of a steel L-bracket, producing a 171{,}504-node tetrahedral mesh, seven analyses \rev{comprising three boundary condition hypotheses together with four load-, material-, and mesh-sensitivity variants}, and a code-compliant assessment revealing structural failure with a quantified redesign. All results are presented as generated in the first autonomous iteration without manual correction\rev{, including the raw residuals between agent-extracted geometry and the physically measured values we supplied afterwards in the supervising-engineer role}, reinforcing that a professional engineer must review and sign off on any such analysis\rev{, and that the resulting residuals are the feedback signal that drives the agent-update operator described in the paper, which we demonstrate only for a single ($k{=}1$) cycle}.
\end{abstract}


\vspace{1em}
\noindent\textbf{Highlights:}
\begin{itemize}[nosep,leftmargin=*]
\item \rev{Photograph $\to$ 171k-node mesh $\to$ seven FEA runs $\to$ code-compliant report in 22 min / \$1}
\item \rev{Solver-agnostic multi-agent architecture with quality gates for FEM, SPH, MPM, DEM, FVM}
\item \rev{Task-dependent conservatism resolves which parameter bound is safe for each limit state}
\item \rev{Image-extracted uncertainty bands validated against physical measurements; residuals reported}
\item \rev{Professional engineer review shown as irreplaceable; single feedback cycle demonstrated}
\end{itemize}

\vspace{1em}
\section{Motivation}
\label{sec:motivation}

\subsection*{Autonomous computational modelling from perceptual data}

We define \emph{autonomous computational modelling from perceptual data} as the capacity of a computational system to execute the complete modelling workflow, from multimodal perceptual data (image, video, audio, text) through inference, modelling, simulation, verification, interpretation, and recommendation, without human intervention at intermediate steps, while maintaining full traceability and requiring professional sign-off on the final output. It is distinct from both \emph{automated simulation} (which executes pre-defined workflows without judgement) and \emph{AI-assisted design} (which augments human decisions on individual steps): the system must \emph{infer} missing information, \emph{apply} engineering judgement to resolve ambiguity, and \emph{recommend} actionable outcomes, not merely compute.

This work sits at the intersection of three converging capabilities: multi-modal large language models that interpret perceptual data and reason about engineering context, multi-agent architectures that decompose complex workflows into specialised communicating roles, and computational mechanics solvers. The contribution is the orchestration layer that connects these into a coherent autonomous modelling pipeline.

\subsection*{The bottleneck}

The bottleneck in modern computational mechanics is rarely the solver (whether finite element, finite volume, particle, or lattice method) but the \emph{human} workflow surrounding it: extracting geometry, selecting boundary conditions, identifying design codes, choosing discretisation strategies, performing verification calculations, interpreting results, \emph{and making recommendations based on those results}. This workflow requires both qualitative judgement (what material is this? what are the likely loading conditions?) and quantitative reasoning (what stress concentration factor is appropriate? which boundary condition variant is conservative for which limit state?). At every stage, the engineer infers as much as possible from the available information to reduce uncertainty, and the framework presented here aims to replicate this inference process autonomously, while making every decision transparent and traceable.

Large language models (LLMs) now demonstrate code generation \cite{chen2021evaluating, li2023starcoder}, scientific reasoning \cite{taylor2022galactica, lewkowycz2022minerva}, and multi-modal understanding \cite{openai2023gpt4}. In computational mechanics specifically: constitutive discovery \cite{linka2023constitutive}, finite element analysis (FEA) code generation \cite{buehler2024mechgpt}, next-generation computer-aided engineering (CAE) \cite{guo2026llmcae}, end-to-end FEA \cite{qi2025feagpt, allfem2026}. Physics-informed neural networks (PINNs) \cite{raissi2019physics, karniadakis2021physics}, neural operators \cite{lu2021deeponet, li2021fourier}, and constitutive artificial neural networks (CANNs) \cite{linka2023constitutive} engage with the mathematical structure of mechanics, while image-based FE modelling \cite{keyak1990automated, hollister1994computational} reconstructs geometry from structured imaging. But each addresses \emph{individual} pipeline components. None orchestrates the \emph{entire} workflow, from observation through inference, simulation, verification, interpretation, and recommendation, which requires \emph{engineering judgement} under uncertainty.

Multi-agent systems \cite{wooldridge2009introduction, dorri2018multi} decompose this orchestration into specialised agents mirroring engineering consultancy practice. MetaGPT \cite{hong2024metagpt}, AutoGen \cite{wu2023autogen}, ReAct \cite{yao2023react}, and Reflexion \cite{shinn2023reflexion} show coordinated LLM agents solving complex tasks; in science, agents have run chemical experiments \cite{boiko2023autonomous} and designed syntheses \cite{bran2024chemcrow}. \rev{In computational mechanics specifically, MechAgents~\cite{nibuehler2024mechagents} demonstrates that a multi-agent LLM team can plan, formulate, code, execute, and critique elasticity FEM problems; the framework presented here is distinguished from that work by starting from perceptual (image) input rather than textual problem statements, and by taking the pipeline through code-compliant assessment and redesign.} We extend this to computational mechanics, where stakes are governed by safety codes and professional liability.

The key contributions are: (i)~a solver-agnostic multi-agent architecture in which agents apply both qualitative and quantitative engineering judgement at every stage, inferring material properties, boundary conditions, and design codes from visual and contextual evidence to maximally reduce uncertainty (\Cref{sec:framework}); (ii)~a layered information extraction and encoding framework with multiple geometry reconstruction pathways and uncertainty quantification, including detection of secondary features such as countersinks from visual cues (\Cref{sec:image_uncertainty}); (iii)~unified mathematical foundations for the finite element method (FEM), smoothed particle hydrodynamics (SPH), material point method (MPM), peridynamics, discrete element method (DEM), and finite volume method (FVM) (\Cref{sec:math}); (iv)~\emph{task-dependent conservatism} resolving the ambiguity of parametric safety analysis (\Cref{sec:conservatism}); (v)~an FEA demonstration where agents not only analyse but \emph{interpret} results and provide actionable engineering recommendations, including autonomous redesign when the component fails (\Cref{sec:results,sec:engineer_role}); and (vi)~a complete reproducible orchestrator prompt (supplied as supplementary material upon acceptance).

\section{General framework}
\label{sec:framework}

\subsection{Architecture overview}
\label{sec:architecture}

\begin{figure}[htbp]
\centering
\includegraphics[width=0.95\textwidth]{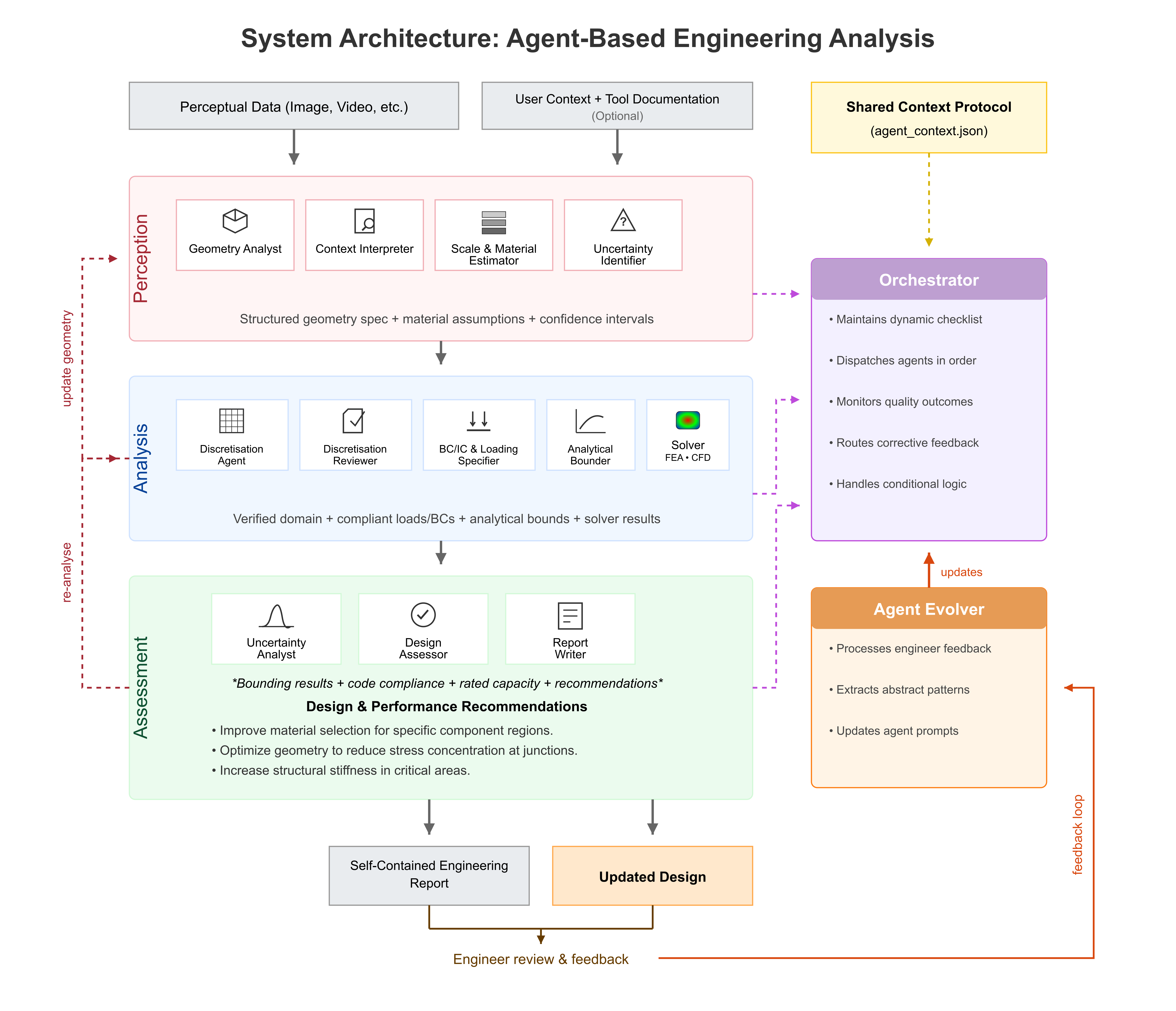}
\caption{Three-layer architecture with orchestrator and iterative feedback. The \emph{input} (perceptual data: image, video, audio, or text, plus user metadata) feeds into three processing layers: the \emph{perception layer} extracts structured data from the perceptual input; the \emph{analysis layer} handles discretisation, solving, and verification; the \emph{assessment layer} interprets results against design codes and emits actionable \emph{recommendations} and proposed \emph{design modifications} as outputs. An overarching \emph{orchestrator} (right) coordinates the entire process, maintaining a dynamic checklist of tasks, dispatching agents, monitoring quality gate outcomes, and routing feedback between layers. The \emph{agent evolver} (top right) processes engineer corrections after each completed analysis, updating agent definitions and shared memory for future runs. Dashed arrows indicate iterative feedback: assessment findings can trigger re-analysis with modified assumptions, and analysis-layer quality gates can request updated perception-layer outputs. The orchestrator decides \emph{which} agent to invoke next based on the current state of the shared context, the pending checklist items, and any gate failures. The architecture is solver-agnostic (FEA, computational fluid dynamics (CFD), DEM, SPH, MPM, lattice Boltzmann method (LBM)).}
\label{fig:general_framework}
\end{figure}

The framework consists of three layers operating in an iterative loop (\Cref{fig:general_framework}). The \emph{perception layer} extracts structured engineering information from unstructured inputs (images, videos, text descriptions). The \emph{analysis layer} executes the computational mechanics workflow: discretisation (meshing, particle seeding, or lattice generation), solver execution, and verification. The \emph{assessment layer} interprets results in the context of design codes and delivers actionable engineering output. Critically, information flows not only forward but also backward: the assessment layer may flag issues requiring re-analysis with modified assumptions, and the analysis layer may request updated geometry or boundary conditions from the perception layer. Quality gates at layer boundaries enforce this iterative refinement.

Each agent within the framework is characterised by four properties. First, a \emph{scope definition} that precisely delimits what the agent is responsible for. Second, a \emph{tool set} specifying which computational tools the agent may invoke. Third, a \emph{critic protocol} requiring the agent to self-validate its outputs. Fourth, a \emph{memory interface} allowing the agent to read lessons from previous runs and write new lessons for future improvement. Agents communicate through structured JSON via a shared context protocol, not natural language, reducing the risk of information loss or misinterpretation across the pipeline.

Two additional meta-agents sit outside the pipeline layers. The \emph{orchestrator} $\mathcal{O}$ coordinates the entire process: it maintains a dynamic checklist $\mathcal{L} = \{(\ell_k, \mathrm{status}_k)\}$ of pending tasks, dispatches agents in the correct order, monitors quality gate outcomes, and routes corrective feedback between layers when a gate fails. When a discretisation reviewer returns $\mathcal{G}_j = 0$, for example, the orchestrator appends corrective instructions to the context and re-invokes the discretisation generator with the updated context. The orchestrator also handles conditional logic: if the problem formulator determines that dynamic analysis is required, the orchestrator inserts the appropriate solver invocations into the execution plan. The number of agents $n$ is not fixed: \Cref{fig:general_framework} shows the full catalogue of 12 specialised agent types across the three layers (4 perception, 5 analysis, 3 assessment), but the orchestrator instantiates and may merge roles as needed based on the problem type. For example, in the FEA demonstration (\Cref{sec:results}), the four perception-layer roles (geometry analyst, context interpreter, scale \& material estimator, uncertainty identifier) are handled by a single geometry analyst agent, and the convergence assessor and bound verifier roles are absorbed by the uncertainty analyst, yielding nine instantiated agents from the twelve defined types. The \emph{agent evolver} $\mathcal{F}$ (\Cref{eq:feedback}) operates after pipeline completion, processing engineer feedback to update agent definitions for future runs.

The same framework orchestrates FEA stress analysis, CFD flow simulation, or DEM granular flow, with only the discretisation and solver agents requiring specialisation.

\subsection{Mathematical formalisation}
\label{sec:formal}

Each agent is instantiated on Claude Opus~4.6 \cite{anthropic2025claude}, a multimodal autoregressive language model (LM) \rev{\cite{vaswani2017attention}} with context window $C \approx 10^6$ tokens, defining
\begin{equation}
    p_{\mathrm{LM}}(y_{1:T} \mid x_{1:S}) = \prod_{t=1}^{T} p_{\mathrm{LM}}(y_t \mid x_{1:S}, y_{1:t-1}).
    \label{eq:autoregressive}
\end{equation}
\rev{Here $x_{1:S} = (x_1, \ldots, x_S)$ is the input token sequence of length $S$ and $y_{1:T} = (y_1, \ldots, y_T)$ is the output token sequence of length~$T$.}
An \emph{agent} $\mathcal{A}_i$ is defined by four components: a system prompt $s_i \in \mathcal{V}^*$ encoding its scope, responsibilities, and domain knowledge; a tool set $\mathcal{T}_i = \{t_1, \ldots, t_{K_i}\}$ of functions (file I/O, shell commands, solvers, or other agents) the agent may invoke; each tool may be deterministic, stochastic, or a hybrid combination (e.g., a deterministic solver wrapped by an LLM-driven input generator), so the agent-tool system is in general a hybrid stochastic-deterministic one; a persistent memory $m_i \in \mathcal{M}$ storing lessons from previous runs; and the underlying language model $p_{\mathrm{LM}}$. The agent operator is
\begin{equation}
    \mathcal{A}_i \equiv \mathcal{A}(s_i, \mathcal{T}_i, m_i, p_{\mathrm{LM}}) : \mathcal{X}_i \times \mathcal{C} \to \mathcal{Y}_i \times \mathcal{C},
    \label{eq:agent_operator}
\end{equation}
where $\mathcal{X}_i$ is the input space, $\mathcal{C}$ the shared context space, and $\mathcal{Y}_i$ the output space. The system prompt $s_i$ conditions $p_{\mathrm{LM}}$ to behave as a domain specialist, analogous to boundary conditions in a partial differential equation (PDE): the same model produces fundamentally different behaviour depending on $s_i$. When $p_{\mathrm{LM}}$ generates a tool-call token sequence, the framework executes the corresponding $t \in \mathcal{T}_i$ and feeds the result back as additional context, creating a hybrid stochastic-deterministic system. At completion, the agent may update its memory: $m_i \leftarrow m_i \oplus \Delta m_i$, where $\oplus$ denotes the append operator (concatenation of structured records; it is not modular addition; memory and context are monotonically non-decreasing).

Two families of maps mediate between the physical space $\mathcal{P}$ (perceptual data: images, video frames, audio, text), token space $\mathcal{V}^*$ (LM native), \rev{the} engineering space $\mathcal{E}$ (structured numerical data)\rev{, and the report space $\mathcal{R}$ (structured engineering assessment output, including recommendations and design modifications)}:
\begin{align}
    \mathrm{enc}_{\mathcal{P}} &: \mathcal{P} \to \mathcal{V}^*, \quad
    \mathrm{enc}_{\mathcal{E}} : \mathcal{E} \to \mathcal{V}^*, \label{eq:enc} \\
    \mathrm{dec}_{\mathcal{E}} &: \mathcal{V}^* \to \mathcal{E}, \quad
    \mathrm{dec}_{\mathcal{R}} : \mathcal{V}^* \to \mathcal{R}. \label{eq:dec}
\end{align}
The physical encoder $\mathrm{enc}_{\mathcal{P}}$ is the physical data transformer (for images, the vision transformer component; analogous modality-specific encoders handle video, audio, or text) of the multimodal LM; the engineering encoder $\mathrm{enc}_{\mathcal{E}}$ serialises structured data (JSON, solver output) into tokens. The decoder $\mathrm{dec}_{\mathcal{E}}$ parses structured output from token sequences; this is where many agent errors originate (malformed JSON, unit mismatches, hallucinated values), motivating the quality gates of \Cref{sec:quality}.

The orchestrator $\mathcal{O}$ manages the pipeline execution. It maintains a checklist $\mathcal{L} = \{(\ell_k, \mathrm{status}_k)\}_{k=1}^K$ and selects the next agent to invoke based on the current context $c_i$ and gate outcomes. The nominal pipeline is
\begin{equation}
    \Pi = \mathcal{O}\!\left(\mathcal{A}_1, \ldots, \mathcal{A}_n, \mathcal{G}_1, \ldots, \mathcal{G}_{n-1}\right),
    \label{eq:pipeline}
\end{equation}
where $\mathcal{G}_j : \mathcal{Y}_j \times \mathcal{C} \to \{0, \rev{\tfrac{1}{2}}, 1\}$ are quality gates \rev{with three admissible outcomes: reject ($\mathcal{G}_j = 0$, which triggers corrective iteration), pass with a logged warning ($\mathcal{G}_j = \tfrac{1}{2}$, the ``conditional pass'' state used in \Cref{sec:results} to capture, for example, a mesh-convergence indicator that is marginal but not outright failing, or an analytical-bound check that holds in the bulk of the domain but tolerates a known stress singularity), and pass ($\mathcal{G}_j = 1$). Conditional-pass outcomes do not trigger an automatic re-invocation, but they are written into the shared context $c_j$ as warnings that the supervising engineer must resolve at sign-off time (\Cref{sec:engineer_role}). Formally, pipeline success requires $\min_j \mathcal{G}_j \geq \tfrac{1}{2}$, and unconditional success requires $\min_j \mathcal{G}_j = 1$.} When $\mathcal{G}_j = 0$, $\mathcal{O}$ appends corrective instructions to $c_j$ and re-invokes $\mathcal{A}_j$ (or an earlier agent if the failure requires upstream changes). Context evolves as
\begin{equation}
    c_i = c_{i-1} \oplus \mathcal{A}_i(x_i, c_{i-1})\big|_{\mathcal{C}}, \qquad i = 1, \ldots, n,
    \label{eq:context_evolution}
\end{equation}
with $\oplus$ the append operator introduced above, ensuring the context is monotonically non-decreasing and hence fully traceable. Each agent sees the outputs of \emph{all} predecessors through $c_{i-1}$, enabling long-range dependencies (e.g., the uncertainty analyst references assumptions made by the geometry analyst).

The self-improving architecture is formalised as a feedback operator
\begin{equation}
    \mathcal{F} : (r, \delta r, \{s_i, \mathcal{T}_i, m_i\}_{i=1}^n) \mapsto \{s_i', m_i'\}_{i=1}^n,
    \label{eq:feedback}
\end{equation}
\rev{where $\delta r$ is the \emph{engineer-supplied} correction on the report $r$ produced by the pipeline. Crucially, $\delta r$ is not generated autonomously by the agents: it is the ground-truth signal a qualified professional engineer contributes at review time (for example, by measuring the component physically and marking which geometry entries in \Cref{tab:geometry_validation} need adjustment, or by flagging a boundary-condition choice as unphysical). The operator $\mathcal{F}$ is implemented in this work as a deterministic prompt- and memory-level update: each residual entry $(\hat{\theta}_j, \theta_j^{\mathrm{true}})$ marked by the engineer is appended to the relevant agent's persistent memory $m_i$ as an explicit rule of the form ``for inputs resembling this one, prefer $\theta_j^{\mathrm{true}}$ over $\hat{\theta}_j$'', and any recurring theme across corrections is appended to the agent's system prompt $s_i$ as an additional instruction. No gradient step is taken and $p_{\mathrm{LM}}$ is not modified. The present paper demonstrates only a single iteration ($k{=}1$) of the loop: the autonomous first-iteration output of \Cref{sec:results} is the input to $\mathcal{F}$, and the measured ground truth in \Cref{tab:geometry_validation} is the residual $\delta r^{(1)}$ that the engineer would supply. Convergence over $K > 1$ cycles is neither claimed nor demonstrated here and is explicitly left as future work.} Unlike fine-tuning or reinforcement learning from human feedback (RLHF), \rev{$\mathcal{F}$ as used here is} lightweight, interpretable, and reversible\rev{; when the engineer-supplied corrections on a fixed analysis class saturate the prompt/memory capacity in future work, escalation to fine-tuning or preference optimisation of $p_{\mathrm{LM}}$ becomes a separate, opt-in second tier, which this paper does not execute}.

The computational cost is dominated by LLM inference:
\begin{equation}
    \mathrm{Cost}(\Pi) = \sum_{i=1}^{n} \left( T_i^{\mathrm{in}} \cdot c_{\mathrm{in}} + T_i^{\mathrm{out}} \cdot c_{\mathrm{out}} + \sum_{j=1}^{J_i} \mathrm{Cost}(t_{i,j}) \right),
    \label{eq:cost}
\end{equation}
where $T_i^{\mathrm{in/out}}$ are token counts, $c_{\mathrm{in/out}}$ per-token costs, \rev{$J_i = |\mathcal{T}_i|$ is the number of tool invocations by agent~$i$,} and $\mathrm{Cost}(t_{i,j})$ the wall-clock cost of executing tool $j$ (e.g., running a solver).

\subsection{Quality gates and verification}
\label{sec:quality}

No agent trusts its predecessor unconditionally. The architecture pairs each generative agent with an independent reviewing agent, forming generator-critic dyads reminiscent of the generator-discriminator pairing in generative adversarial networks \cite{goodfellow2014gan}: one agent produces a candidate artefact (mesh, bound estimate, code check), and a second agent, conditioned by a different system prompt and with access to the same evidence, audits it and either accepts or rejects. Unlike a GAN, the two agents are not trained against each other; they share the same underlying $p_{\mathrm{LM}}$, but the prompt-level separation of roles is sufficient to expose errors that a single agent would miss. Four such gates enforce verification \cite{oberkampf2010verification, roache1998verification}\rev{~\cite{royoberkampf2011vvuq,asmevv10}}:

The first gate is \textbf{discretisation review}. An independent reviewer agent analyses the generated computational domain against the original image, counting geometric features independently rather than simply checking the modeller's claimed counts. The reviewer agent also performs \emph{junction connectivity} verification, ensuring that at each feature junction (arm-bend, arm-arm), the mesh is continuous with shared nodes at interfaces, and \emph{profile tracing}, tracing the outside and inside profiles through the mesh node coordinates to check for steps, gaps, or reversals that indicate geometry construction errors.

The second gate is \textbf{mesh-convergence assessment}. A convergence agent re-runs the critical load case on a coarsened mesh, compares peak stress and tip deflection against the fine-mesh solution \rev{\cite{zienkiewicz1987}}, and rejects the discretisation when the relative change exceeds a configurable tolerance; it also screens FEA response quantities against the analytical envelopes produced in the next step as a sanity check before assessment.

The third gate is \textbf{analytical bound verification}. Before accepting solver results downstream, the framework requires that all computed response quantities fall within analytically derived upper and lower bounds \cite{pilkey2008petersons}.

The fourth gate is \textbf{code-compliance assessment}. The design assessment explicitly checks each limit state against the applicable code clause, using the appropriate conservative bound for each check type.

A gate failure triggers iteration back to the responsible agent with corrective context appended, implementing the inter-layer feedback arrows in \Cref{fig:general_framework}. The maximum number of iterations per gate before escalation to human review is a configurable parameter $N_{\max}$ (set to $N_{\max} = 5$ for the demonstration in \Cref{sec:results}).

The agents within each layer are not fixed; they are instantiated by the orchestrator based on the problem type. The \emph{perception layer} contains agents for geometry extraction (with manufacturing-step descriptions and junction connectivity specifications), material inference, and context interpretation. The \emph{analysis layer} contains agents for problem formulation, discretisation method selection, discretisation generation (with feature-relative placement and junction verification), discretisation review (quality gate), boundary/initial condition specification, analytical bounding, solver execution, and convergence assessment. The \emph{assessment layer} contains agents for uncertainty analysis, bound verification (quality gate), design assessment / code compliance (quality gate), and report writing. \Cref{sec:results} demonstrates the framework with an FEA instantiation.

\section{Information extraction and encoding across pipeline layers}
\label{sec:image_uncertainty}

The pipeline transforms unstructured multi-modal perception data into a verified engineering report through successive encoding stages. Each layer in \Cref{fig:general_framework} receives information from its predecessor, extracts and encodes domain-specific content, and passes structured output downstream. We formalise the encoding at each stage.

\subsection{Input: image and context acquisition}
\label{sec:input_encoding}

The pipeline accepts multi-modal perceptual input: images, video frames, audio clips, or textual descriptions, together with optional contextual metadata $\mathcal{M}_{\mathrm{ctx}}$ (user-provided descriptions, application domain, geographic region). In this work we focus on the image modality as the primary use case; the generalisation to video (temporal stacks), audio (e.g., impact or modal signatures), and text is direct through modality-specific encoders of the multimodal LM. For the image case, the input is a pixel array $\mathbf{I} \in \mathbb{R}^{H \times W \times 3}$, which may be a clean engineering drawing, a photograph of an installed component, or a hand-sketch. The framework supports multiple pathways from image to 3D geometry, automatically selecting the most appropriate:

\begin{enumerate}[label=(\roman*),nosep]
    \item \textbf{\rev{Vision-language model (VLM)} direct interpretation.} A \rev{VLM} interprets the image and generates a parametric geometry description as structured JSON. This is the most general pathway, handling arbitrary real-world photographs, but is the most lossy.
    \item \textbf{VLM $\to$ script $\to$ \rev{computer-aided design (CAD)}.} The VLM generates a Python script (using Gmsh, OpenCASCADE, or CadQuery) that programmatically constructs the 3D geometry. This combines the VLM's interpretation ability with exact parametric geometry.
    \item \textbf{Edge detection $\to$ 2D profile $\to$ extrusion.} Canny/Hough transforms extract 2D geometric primitives (lines, arcs, circles), which are extruded or revolved to create 3D geometry. Effective for prismatic or axisymmetric components.
    \item \textbf{Segmentation $\to$ point cloud $\to$ surface reconstruction.} Segment Anything Model \rev{\cite{kirillov2023sam}} or Mask R-CNN isolates the component, depth estimation (monocular or stereo) produces a point cloud, and Poisson surface reconstruction generates a watertight mesh.
    \item \textbf{Multi-view $\to$ structure from motion (SfM) $\to$ mesh.} When multiple views are available, SfM algorithms (e.g., COLMAP \cite{schoenberger2016sfm}) produce a dense 3D point cloud that is meshed directly.
    \item \textbf{Text description $\to$ parametric CAD.} When the image is supplemented by textual specifications, the framework bypasses image geometry extraction and constructs the CAD model from the specification directly.
\end{enumerate}

The orchestrator selects the pathway based on image quality, component complexity, and available metadata.

\subsection{Perception layer: geometry and material encoding under uncertainty}
\label{sec:perception_encoding}

The perception layer encodes the input into structured engineering parameters. The geometry agent produces a \emph{manufacturing-step description}: base stock definition, sequential operations (bends with inner/outer radii, holes with edge features, cuts), each located in an explicit coordinate system with \emph{junction connectivity} requirements specifying which surfaces of adjacent features must be flush. Dimensions carry uncertainty bounds $[\theta^-, \theta^+]$, hot-spots are predicted with $K_t$ estimates, and iterative sanity checks verify geometric consistency (bend radius $\geq$ thickness, holes within members, developed length). Full agent instructions will be released with the supplementary material.

Let $\boldsymbol{\theta}_{\mathrm{img}} = (\theta_1, \ldots, \theta_q) \in \mathbb{R}^q$ denote the geometric and material parameters to be extracted. The image extraction operator maps the input to a set-valued estimate:
\begin{equation}\label{eq:image_extraction}
    \mathcal{I} : (\mathbf{I}, \mathcal{M}_{\mathrm{ctx}}) \mapsto \bigl\{(\hat{\theta}_j, \theta_j^-, \theta_j^+, p_j, \kappa_j)\bigr\}_{j=1}^q,
\end{equation}
where $\hat{\theta}_j$ is the point estimate, $[\theta_j^-, \theta_j^+]$ the plausible interval, $p_j(\theta_j \mid \mathbf{I})$ a conditional density, and $\kappa_j \in [0,1]$ a confidence score.

\paragraph{Geometric parameters} For dimensions without calibration \rev{\cite{moore1966interval}}:
\begin{equation}\label{eq:interval_dim}
    L = \hat{L} \pm \delta L, \quad H = \hat{H} \pm \delta H, \quad t = \hat{t} \pm \delta t,
\end{equation}
with $\delta$ depending on scale reference quality ($\pm 5\%$ with calibration object, $\pm 10$ to $20\%$ from contextual references). \rev{The bands $\delta L$, $\delta H$, $\delta t$ are the agent's \emph{self-reported plausibility tags} from image evidence, not probabilistically calibrated confidence intervals: because no ground truth is available from the image alone, the bands cannot carry an empirically verified coverage probability, and should be read as structured expressions of the agent's own confidence that the supervising engineer (\Cref{sec:engineer_role}) compares against physical measurement. The residual between agent estimate and engineer-supplied ground truth is the feedback signal $\delta r$ driving the update operator~\Cref{eq:feedback}.} The density representation enables richer modelling:
\begin{equation}\label{eq:density_params}
    \theta_j \sim
    \begin{cases}
        \mathcal{N}(\hat{\theta}_j, \sigma_j^2), & \text{measurement-derived}, \\
        \mathcal{U}(\theta_j^-, \theta_j^+), & \text{only bounds known}, \\
        \mathrm{Tri}(\theta_j^-, \hat{\theta}_j, \theta_j^+), & \text{educated guess with best estimate}.
    \end{cases}
\end{equation}
The triangular distribution is appropriate for image-inferred parameters, encoding bounds and mode without spurious tails.

\paragraph{Material identification} For qualitative parameters (material type, surface finish, connection type), the extraction produces \emph{fuzzy} membership functions \rev{in the sense of Zadeh~\cite{zadeh1965fuzzy}} $\mu_j : \Theta_j \to [0,1]$:
\begin{equation}\label{eq:fuzzy_membership}
    \mu_{\mathrm{material}}(\theta) =
    \begin{cases}
        \mu_1, & \theta = \text{hypothesis 1}, \\
        \mu_2, & \theta = \text{hypothesis 2}, \\
        \mu_3, & \theta = \text{hypothesis 3},
    \end{cases}
\end{equation}
where $\sum_k \mu_k = 1$ and the values are inferred from visual cues (surface texture, colour, reflectivity), section profile, and application context. When the dominant hypothesis exceeds $\mu_{\min} \approx 0.6$, the framework proceeds and flags the assumption; otherwise it analyses multiple hypotheses. Material properties are marginalised over this uncertainty:
\begin{equation}\label{eq:material_marginal}
    p(E \mid \mathbf{I}) = \sum_{k} \mu_k \, p(E \mid \theta_k),
\end{equation}
yielding mixture distributions that propagate compounded identification and property variability. \rev{We note that when $\sum_k \mu_k = 1$ and the $\mu_k$ are non-negative, the expression is formally identical to a Bayesian model average with prior weights $P(M_k \mid \mathbf{I}) = \mu_k$; we adopt this probabilistic interpretation throughout the downstream analysis, so that $p(E \mid \mathbf{I})$ is a valid probability density.}

\paragraph{Information content} The information gain quantifies extraction quality:
\begin{equation}\label{eq:info_gain}
    \Delta H = H(\boldsymbol{\theta}_{\mathrm{img}}) - H(\boldsymbol{\theta}_{\mathrm{img}} \mid \mathbf{I}) \geq 0,
\end{equation}
where $H$ denotes Shannon entropy. Parameters with high residual entropy $H(\theta_j \mid \mathbf{I})$ are candidates for targeted physical measurement, providing a principled basis for requesting additional data.

\subsection{Analysis layer: encoding into computational models}
\label{sec:analysis_encoding}

The analysis layer receives the structured perception output and encodes it into computational models. This involves three encoding decisions:

\paragraph{Problem formulation encoding} The problem formulator determines the governing physics (structural, fluid, thermal, coupled), selects the strong form PDE (\Cref{sec:math}), and classifies the loading type (load-controlled vs.\ displacement-controlled), which governs the conservatism framework (\Cref{sec:conservatism}).

\paragraph{Discretisation encoding} The discretisation selector chooses the method based on problem type: structural problems without fracture use FEM (CalculiX or FEniCS); fracture problems use peridynamics (Peridigm); large-deformation problems with history-dependent materials use MPM (CB-Geo); granular flows use DEM (YADE, MercuryDPM, or LIGGGHTS); free-surface fluids use SPH (DualSPHysics); enclosed fluids use FVM (OpenFOAM); thermal problems use FEM or FVM; and coupled problems use partitioned solvers with preCICE. The generator constructs geometry using \emph{feature-relative placement}: computing junction coordinates analytically from the bend geometry, placing each arm flush with the fillet endpoints, and verifying a single connected solid after boolean fuse. This prevents the most common error of placing features at absolute coordinates without verifying surface coincidence. Visual comparison images are generated for the reviewer.

\paragraph{Boundary condition and loading encoding} Boundary conditions (BCs), initial conditions (ICs), and loads are encoded from the image and context. The applicable design code is \emph{inferred} (not prescribed) from geographic clues, material type, and application. Multiple BC variants are generated, each tagged with conservatism direction per limit state.

\subsection{Assessment layer: encoding into engineering judgement}
\label{sec:assessment_encoding}

The assessment layer encodes numerical results into engineering decisions: the \emph{uncertainty analyst} applies task-dependent conservatism (\Cref{sec:conservatism}) producing separate \rev{factor of safety (FoS)} ranges per limit state; the \emph{bound verifier} (quality gate) checks results against analytical bounds, diagnosing exceedances; the \emph{design assessor} performs code-compliant checks and, critically, \emph{interprets} results to provide actionable recommendations, invoking a redesign loop when the component fails; and the \emph{report writer} produces a professional simulation qualification report with full reproducibility.

\section{Mathematical foundations for solver-agnostic discretisation}
\label{sec:math}

\subsection{The forward problem: continuum fields and what must be known}

Engineering simulation is a \emph{forward problem}: given geometry $\Omega$, material constitutive law $\bsig = \hat{\bsig}(\beps, \dot{\beps}, \boldsymbol{\alpha}, T, \ldots)$, boundary conditions on $\partial\Omega$, initial conditions at $t=0$ (for dynamic problems), and applied loads, find the response fields (displacements, stresses, velocities, pressures, temperatures). \Cref{tab:continuum_physics} enumerates the continuum field equations for the three classical domains (solid, fluid, and granular), identifying the unknown fields, the number of scalar equations, the constitutive closure required, and the additional information that must be supplied to render the problem complete.

\begin{table}[htbp]
\centering
\caption{Continuum physics: fields, equations, and required inputs. This table is not exhaustive; many other physics models (piezoelectric, magnetohydrodynamic, phase-field fracture, etc.) follow the same pattern. For each row, agents must determine \emph{all} listed quantities before a solver can be invoked.}
\label{tab:continuum_physics}
\resizebox{\textwidth}{!}{%
\footnotesize
\begin{tabular}{@{}lcccccl@{}}
\toprule
\textbf{Physics} & \textbf{Fields} & \textbf{Scalar eqns} & \textbf{PDE type} & \textbf{ICs?} & \textbf{State?} & \textbf{Constitutive + additional inputs} \\
\midrule
\multicolumn{7}{l}{\textit{Solid mechanics}} \\
Static elastic & $\bu$ & $d$ & Elliptic & No & No & $E,\nu$ (or $\bC$); BCs ($\bar{\bu},\bar{\bt}$); $\bb$ \\
Dynamic elastic & $\bu$, $\dot{\bu}$ & $d$ & Hyperbolic & Yes & No$^\dagger$ & $E,\nu,\rho$; BCs; $\bu_0,\dot{\bu}_0$ \\
Elastoplastic & $\bu$, $\boldsymbol{\alpha}$ & $d{+}n_\alpha$ & Hyp.\ + evol. & Yes & Yes & $E,\nu$, yield $f$, hardening $H$; $\boldsymbol{\alpha}_0$ \\
Hyperelastic (large def.) & $\bu$, $\bF$ & $d$ & Elliptic (NL) & No & No & $W(\bF)$ strain energy; BCs; $\bb$ \\
\midrule
\multicolumn{7}{l}{\textit{Fluid mechanics}} \\
Incompressible & $\bv$, $p$ & $d{+}1$ & Parabolic+constr. & Yes & No & $\rho,\mu$; BCs ($\bar{\bv},\bar{p}$); $\bv_0$ \\
Compressible & $\rho$, $\bv$, $E$ & $d{+}2$ & Hyperbolic & Yes & No & EOS $p(\rho,E)$, $\mu$, $k$; BCs; ICs \\
\midrule
\multicolumn{7}{l}{\textit{Thermal}} \\
Heat conduction & $T$ & 1 & Parabolic & Yes$^\ddagger$ & No & $k,\rho,c_p$; BCs ($\bar{T},\bar{q}$); $T_0$ \\
\midrule
\multicolumn{7}{l}{\textit{Granular / particulate}} \\
Granular flow & $\bv$, $\bsig$, $\phi$ & $d{+}6{+}1$ & Hyp.\ + evol. & Yes & Yes & $\mu_s,\phi_{\max}$, dilation; $\phi_0$ \\
\midrule
\multicolumn{7}{l}{\textit{Coupled}} \\
Thermo-mechanical & $\bu$, $T$ & $d{+}1$ & Elliptic+parabolic & Yes & Depends & $E(T),\nu,\alpha_T,k,c_p$; BCs; $T_0$ \\
Fluid-structure (FSI) & $\bu_s$, $\bv_f$, $p$ & $2d{+}1$ & Mixed & Yes & Depends & Solid+fluid props; interface conditions \\
Thermofluid & $\bv$, $p$, $T$ & $d{+}2$ & Parabolic & Yes & No & $\rho,\mu,k,c_p,\beta$; BCs; $\bv_0,T_0$ \\
\bottomrule
\multicolumn{7}{l}{\footnotesize $^\dagger$For path-dependent materials (viscoelastic, damage), add state variables $\boldsymbol{\alpha}$ with evolution equations.} \\
\multicolumn{7}{l}{\footnotesize $^\ddagger$Steady-state heat conduction is elliptic (no ICs); transient is parabolic (requires $T_0$).} \\
\multicolumn{7}{l}{\footnotesize Notation: $\bu$ displacement, $\bv$ velocity, $p$ pressure, $T$ temperature, $\bF$ deformation gradient, $\bsig$ stress, $d$ spatial dimension,} \\
\multicolumn{7}{l}{\footnotesize $E$ Young's modulus, $\nu$ Poisson's ratio, $\rho$ density, $\mu$ dynamic viscosity, $k$ thermal conductivity, $c_p$ specific heat,} \\
\multicolumn{7}{l}{\footnotesize $\bb$ body force, $\bar{\bu}$/$\bar{\bt}$ prescribed displacement/traction, $\phi$ solid fraction, $\alpha_T$ thermal expansion, $\beta$ buoyancy coefficient.}
\end{tabular}%
}
\end{table}

Every entry in \Cref{tab:continuum_physics} represents information that the agents must either \emph{extract} from the image (geometry, material appearance), \emph{infer} from engineering judgement (BCs, loads, whether the problem is static or dynamic, whether thermal effects matter), or \emph{select} from domain knowledge (constitutive model, need for initial conditions, presence of internal state). Dynamic problems always require initial conditions; path-dependent materials always carry internal state variables $\boldsymbol{\alpha}$ that must be initialised and evolved; coupled problems require interface conditions between domains. The table is not exhaustive; piezoelectric, magnetohydrodynamic, phase-field fracture, chemically reactive, and many other physics models follow the same pattern of fields, equations, constitutive closure, and boundary/initial data.

\paragraph{Coupled multi-physics} Many engineering problems involve coupling between two or more physics from \Cref{tab:continuum_physics}. Thermo-mechanical coupling requires temperature-dependent material properties and thermal expansion; fluid-structure interaction (FSI) requires kinematic and dynamic conditions at the fluid-solid interface; thermofluid problems couple the Navier-Stokes equations with energy transport. Coupled solvers may handle all physics in a single code (monolithic), or separate solvers may be coupled via libraries such as preCICE \cite{chourdakis2022} using message-passing interface (MPI) data exchange at shared interfaces. The discretisation selector agent must identify whether coupling is required and, if so, whether a monolithic or partitioned approach is appropriate.

\subsection{The inverse problem: agents as conditioned operators}

Before the forward problem can be solved, the agents must address what is fundamentally an \emph{inverse problem}: given a photograph $\mathbf{I}$ and optional context $\mathcal{M}_{\mathrm{ctx}}$, determine the geometry $\Omega$, material parameters $\boldsymbol{\theta}_{\mathrm{mat}}$, boundary conditions, and loading that define the forward problem. This is precisely the image extraction operator of \Cref{eq:image_extraction}. Framing it as an inverse problem clarifies why the agents are effective: each agent $\mathcal{A}_i = \mathcal{A}(s_i, \mathcal{T}_i, m_i, p_{\mathrm{LM}})$ from \Cref{eq:agent_operator} is a \emph{conditioned operator}, conditioned by its system prompt $s_i$ and memory $m_i$ on the accumulated knowledge of engineering practice. The language model $p_{\mathrm{LM}}$, trained on engineering textbooks, design codes, and solver documentation, provides a strong prior for this inverse problem. The agent's output is not a blind guess but a \emph{maximum a~posteriori} estimate conditioned on both the image evidence and the engineering prior, with uncertainty bounds that propagate through the pipeline.

This inverse-problem perspective explains the three-layer architecture. The perception layer solves the \emph{geometric and material inverse}: what component is this, and what is it made of? The analysis layer solves the \emph{forward problem}: given the inferred inputs, what are the response fields? The assessment layer solves the \emph{design inverse}: given the response, does the component satisfy the code requirements, and if not, what changes would make it adequate?

\subsection{From continuum to algebra: discretisation and solvers}

The continuum PDEs of \Cref{tab:continuum_physics} define fields over infinite-dimensional function spaces. Discretisation transforms them into finite-dimensional algebraic systems ($\mathbf{K}\mathbf{U} = \mathbf{F}$ for static problems, $\mathbf{M}\ddot{\mathbf{U}} + \mathbf{C}\dot{\mathbf{U}} + \mathbf{K}\mathbf{U} = \mathbf{F}$ for dynamics) that can be solved by linear algebra routines. The choice of discretisation method determines the structure of these algebraic systems, and the choice of \emph{solver software} determines the available element types, constitutive models, contact formulations, and output formats.

Solvers may be commercial (ABAQUS, ANSYS, COMSOL), open-source (CalculiX \cite{dhondt2004calculix}, OpenFOAM, FEniCS \cite{logg2012}, deal.II \cite{arndt2021}), or in-house research codes. Each solver encodes specific physics capabilities, constitutive model libraries, and input/output formats in its documentation. The solver agent must consult this documentation to assemble valid input files, a task for which LLMs are particularly well-suited, as they can parse and reason about technical documentation. The strong and weak forms presented below are the mathematical content that every solver implements internally; the agent's role is to select the appropriate solver and provide it with the correct inputs, not to re-derive the discretisation.

\subsection{Strong and weak forms}

Whether a method works from the strong form (pointwise PDE satisfied at collocation points or particles, as in SPH, classical DEM, and finite difference / collocation schemes) or from the weak form (integral statement tested against admissible variations, as in FEM, FVM, and MPM) is itself part of the discretisation choice. We state both forms below; individual solvers implement the one appropriate to their method. For solid mechanics, consider a body $\Omega \subset \R^d$ with $\partial\Omega = \Gamma_D \cup \Gamma_N$. The \textbf{strong form} (momentum balance):
\begin{equation}\label{eq:strong}
    \nabla \cdot \bsig + \bb = \rho \ddot{\bu} \quad \text{in } \Omega,
\end{equation}
with constitutive $\bsig = \hat{\bsig}(\beps, \dot{\beps}, \boldsymbol{\alpha})$, BCs $\bu = \bar{\bu}$ on $\Gamma_D$, $\bsig \cdot \bn = \bar{\bt}$ on $\Gamma_N$, and ICs $\bu(\bx,0) = \bu_0$, $\dot{\bu}(\bx,0) = \dot{\bu}_0$ for dynamics. The \textbf{weak form}: find $\bu \in \mathcal{V}$ such that $\forall \bv \in \mathcal{V}_0$\rev{, where $\mathcal{V} = \{\bu \in H^1(\Omega)^d : \bu = \bar{\bu} \text{ on } \Gamma_D\}$ is the space of kinematically admissible displacements and $\mathcal{V}_0 = \{\bv \in H^1(\Omega)^d : \bv = \mathbf{0} \text{ on } \Gamma_D\}$ is its homogeneous counterpart}:
\begin{equation}\label{eq:weak}
    \int_\Omega \bsig : \beps(\bv) \dif\Omega
    = \int_\Omega \bb \cdot \bv \dif\Omega
    + \int_{\Gamma_N} \bar{\bt} \cdot \bv \dif\Gamma
    + \int_\Omega \rho \ddot{\bu} \cdot \bv \dif\Omega.
\end{equation}
For statics with linear elasticity, this reduces to minimising $\Pi(\bu) = \tfrac{1}{2}\int_\Omega \beps : \bC : \beps \dif\Omega - \int_\Omega \bb \cdot \bu \dif\Omega - \int_{\Gamma_N} \bar{\bt} \cdot \bu \dif\Gamma$. Analogous strong/weak forms exist for each physics in \Cref{tab:continuum_physics}; the key point is that every solver implements one or more of these forms internally, and the agent's task is to provide the correct inputs from the rows of \Cref{tab:continuum_physics}.

\subsection{Discretisation methods}

Each method transforms the weak form into an algebraic system differently:

\paragraph{Finite element method (mesh-based)} Decompose $\Omega = \bigcup_e \Omega_e$, approximate $\bu^h = \sum_a N_a \bu_a$, yielding $\mathbf{K}\mathbf{U} = \mathbf{F}$ with element stiffness $\mathbf{K}_e = \int_{\Omega_e} \mathbf{B}^T \mathbf{D} \mathbf{B} \dif\Omega$. Available in CalculiX \cite{dhondt2004calculix}, FEniCS \cite{logg2012}, deal.II \cite{arndt2021}, and commercial codes.

\paragraph{Smoothed particle hydrodynamics (meshfree)} Kernel approximation $\langle f(\bx) \rangle \approx \sum_b (m_b/\rho_b) f(\bx_b) W(\bx - \bx_b, h)$ transforms PDEs into ODE systems on particles \cite{monaghan2005, liu2003}. Available in DualSPHysics \cite{dualsphysics2022}, LAMMPS \cite{thompson2022lammps}.

\paragraph{Material point method (hybrid)} Material points carry state through a background grid \cite{sulsky1994, zhang2017}\rev{~\cite{sulsky1995}}. Grid quantities assembled as $m_I^L = \sum_p N_I(\bx_p) m_p$. Available in CB-Geo, Taichi MPM.

\paragraph{Peridynamics (nonlocal)} Replaces $\nabla \cdot \bsig$ with an integral $\int_{\mathcal{H}_\delta} \bff(\bu(\bx') - \bu(\bx), \bx' - \bx) \dif V_{\bx'}$ \cite{silling2000, silling2007}, naturally handling fracture. Available in Peridigm.

\paragraph{Discrete element method (particle-based, explicit)} Models granular and particulate materials as collections of rigid or deformable particles interacting through contact forces. Newton's second law is integrated for each particle $i$: $m_i \ddot{\bx}_i = \sum_j \bff_{ij}^{\mathrm{contact}} + m_i \mathbf{g}$, where contact forces $\bff_{ij}$ are computed from overlap, relative velocity, and friction models (Hertz-Mindlin, linear spring-dashpot). Naturally handles granular flow, crushing, mixing, and segregation. Available in YADE \cite{smilauer2015yade}, MercuryDPM \cite{weinhart2020mercurydpm}, LIGGGHTS \cite{kloss2012liggghts}, BlazeDEM \cite{govender2018blazedem}, and LAMMPS \cite{thompson2022lammps}.

\paragraph{Finite volume method (conservation-form)} Cell-centred discretisation of $\partial_t \int_{V_i} \mathbf{Q} \dif V + \oint_{\partial V_i} \mathbf{F} \cdot \bn \dif S = \int_{V_i} \mathbf{S} \dif V$ using numerical fluxes. Available in OpenFOAM \cite{weller1998openfoam}, SU2 \cite{economon2016su2}.

\subsection{Discretisation selection and solver documentation}

The preceding subsections focused on structural mechanics because the demonstration in \Cref{sec:results} is a structural problem. The same philosophy applies directly to fluids (FVM discretisation, OpenFOAM solver), heat transfer (FEM with thermal elements, or FVM), granular flows (DEM or MPM), and coupled problems (partitioned or monolithic solvers). In every case, the agent's task is the same: identify the physics from \Cref{tab:continuum_physics}, select the appropriate discretisation, consult the solver documentation to assemble valid input files, and provide the correct fields, constitutive parameters, and boundary/initial conditions.

The discretisation selection logic takes problem type~$\mathcal{P}$, material model~$\mathcal{M}$, expected deformation magnitude~$\delta_{\max}$, and a fracture flag~$f$ as inputs, and returns a discretisation method~$\mathcal{D}$ and solver~$\mathcal{C}$. The selection proceeds as follows. For structural problems: if fracture is expected, peridynamics (Peridigm) is selected; if large deformations exceed a threshold $\delta_{\max} > \delta_{\mathrm{thr}}$ and the material is history-dependent, MPM (CB-Geo) is selected; otherwise FEM (CalculiX or FEniCS) is used. For granular flows, DEM (YADE, MercuryDPM, or LIGGGHTS) is selected. For fluid problems, SPH (DualSPHysics) is chosen when free surfaces are present, and FVM (OpenFOAM) otherwise. Thermal problems use FEM or FVM, and coupled problems use partitioned solvers with preCICE for inter-solver communication.

\section{Task-dependent conservatism}
\label{sec:conservatism}

``What does `conservative' mean?'' The answer depends on the \emph{loading type}, the \emph{limit state}, and the \emph{parameter} being varied. Although we illustrate with structural examples (because the demonstration in \Cref{sec:results} is structural), the principle is general: in CFD, a conservative inlet velocity estimate differs for drag assessment vs.\ lift assessment; in heat transfer, a conservative thermal conductivity differs for maximum-temperature vs.\ thermal-stress predictions.

\paragraph{The core issue} For any response function $f_k(\boldsymbol{\theta})$ depending on uncertain parameters $\boldsymbol{\theta} \in \Theta$, the conservative value is $\max_{\boldsymbol{\theta}} f_k$ if the limit state is exceeded when $f_k$ is \emph{large} (e.g., stress exceeding yield), or $\min_{\boldsymbol{\theta}} f_k$ if the limit state is exceeded when $f_k$ is \emph{small} (e.g., buckling load falling below applied load). Different limit states $k$ may have different maximising/minimising $\boldsymbol{\theta}^*$, making a single ``conservative'' parameter set logically impossible.

\paragraph{Structural example} In load-controlled problems (constant applied force), flexible BCs increase both peak stress and deflection, conservative for both strength and stiffness. In displacement-controlled problems (prescribed displacement), stiffer structures generate higher internal forces; the conservatism direction reverses. Stress concentrations at geometric discontinuities ($K_t$) further complicate the picture, as stiffer structures cannot redistribute load around these features.

\paragraph{Formal framework} Let $\boldsymbol{\theta} \in \Theta \subset \mathbb{R}^p$ with intervals $\theta_j \in [\theta_j^-, \theta_j^+]$ \rev{\cite{ben-tal2009robust}}. The task-dependent conservative values are:
\begin{equation}\label{eq:conservatism}
    f_{k,\mathrm{cons}} = \begin{cases} \max_{\boldsymbol{\theta} \in \Theta} f_k(\boldsymbol{\theta}), & \text{if limit state triggered by large } f_k, \\ \min_{\boldsymbol{\theta} \in \Theta} f_k(\boldsymbol{\theta}), & \text{if limit state triggered by small } f_k. \end{cases}
\end{equation}
Adequacy requires simultaneous satisfaction across all limit states:
\begin{equation}\label{eq:simultaneous}
    \text{PASS} \iff \bigwedge_{k} g_k(f_{k,\mathrm{cons}}) \leq 0.
\end{equation}
In practice, $\Theta$ is explored through $N$ discrete variants, reporting the factor of safety (FoS) range:
\begin{equation}\label{eq:fos_range}
    \mathrm{FoS}_{\min}^{\mathrm{strength}} = \min_i \frac{f_y}{f_s(\boldsymbol{\theta}_i)}, \qquad
    \mathrm{FoS}_{\max}^{\mathrm{strength}} = \max_i \frac{f_y}{f_s(\boldsymbol{\theta}_i)},
\end{equation}
appropriately representing the uncertainty range rather than a single misleading value.

\section{Demonstration: Steel L-bracket analysis}
\label{sec:results}

We demonstrate the framework applied to a zinc-plated steel L-bracket. The \textbf{complete input} to the pipeline is a single photograph of the component held against a wall in its intended installation orientation (\Cref{fig:input_photo}); no CAD files, engineering drawings, material certificates, dimensional measurements, or user text are supplied. Every engineering parameter (geometry, material grade, design code, boundary conditions, \emph{load magnitudes}, and safety factors) is \emph{inferred} by the agents from the image alone. In particular, the service load is inferred by the BC/load-specifier agent from visual cues (shelf-bracket geometry, zinc plating, wall-mount hole pattern) combined with engineering practice for domestic wall-mounted shelving. The orchestrator instantiates nine agents from the twelve types defined in \Cref{sec:framework}: the four perception-layer roles are merged into a single geometry analyst (Agent~1), the five analysis-layer roles map to Agents~2 to 6 (mesh modeller, mesh reviewer, BC/load specifier, analytical bounder, FEA solver), and the three assessment-layer roles map to Agents~7 to 9 (uncertainty analyst, which also absorbs the convergence assessor and bound verifier roles, design recommender, and report writer). For this structural FEA problem, the discretisation selector chooses FEM with Gmsh meshing and CalculiX solving. Each subsection describes the responsible agent's role, the qualitative and quantitative judgements it makes, and its output.

\subsection{Agent 1 (Perception): Geometry analyst}

\textbf{Role.} Receives a photograph and produces a manufacturing-step description: base stock definition, sequential operations (bends, holes) with explicit coordinate system, inner-bend/outer-bend radius identification, and junction connectivity requirements specifying which surfaces of adjacent features must be flush.

\begin{figure}[htbp]
\centering
\begin{subfigure}[b]{0.45\textwidth}
    \centering
    \includegraphics[width=\textwidth]{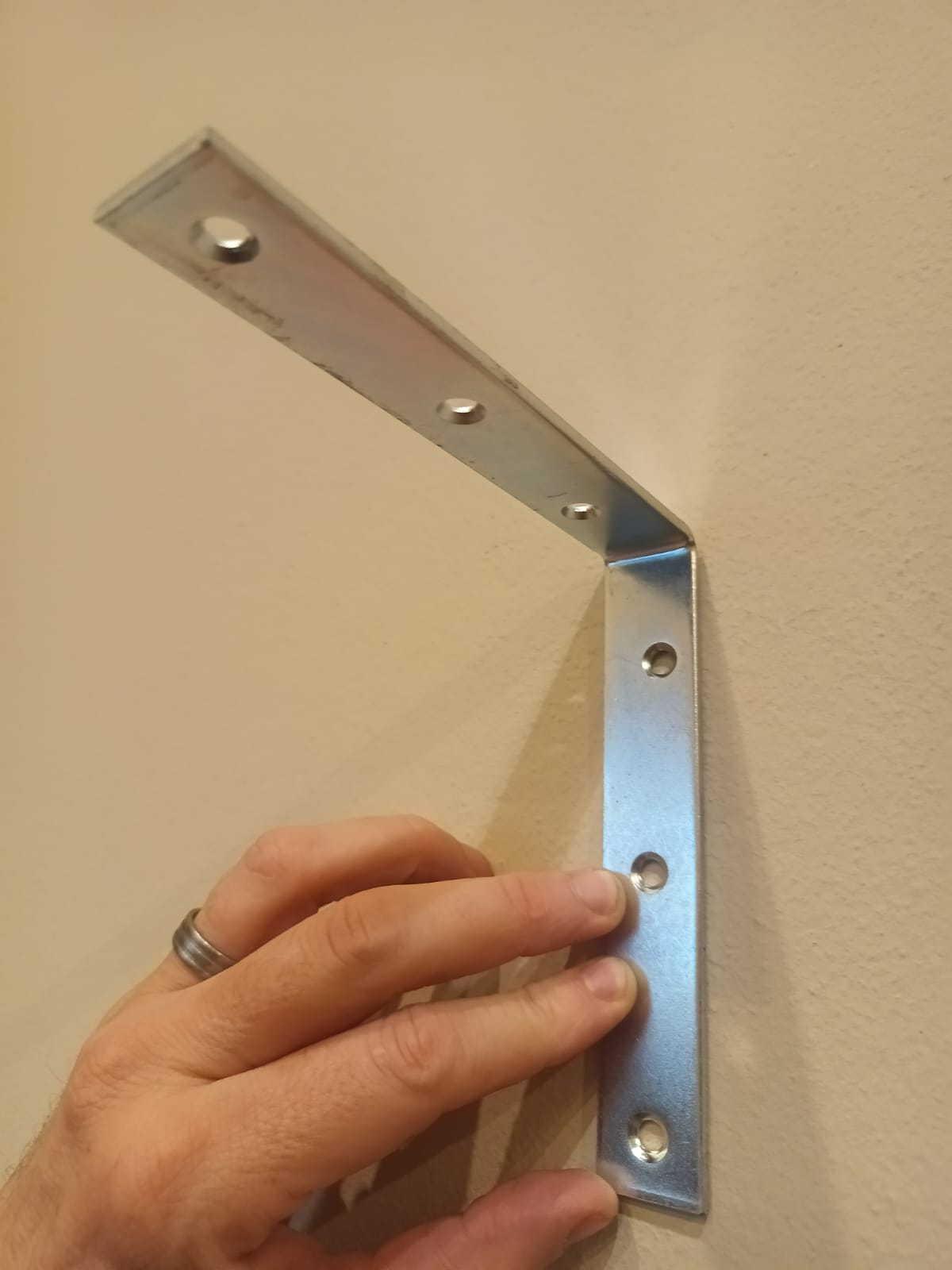}
    \caption{Input photograph. The geometry analyst uses the hand as scale reference and identifies six holes, the $90^\circ$ bend, and zinc plating to infer material grade.}
    \label{fig:input_photo}
\end{subfigure}
\hfill
\begin{subfigure}[b]{0.45\textwidth}
    \centering
    \includegraphics[width=\textwidth]{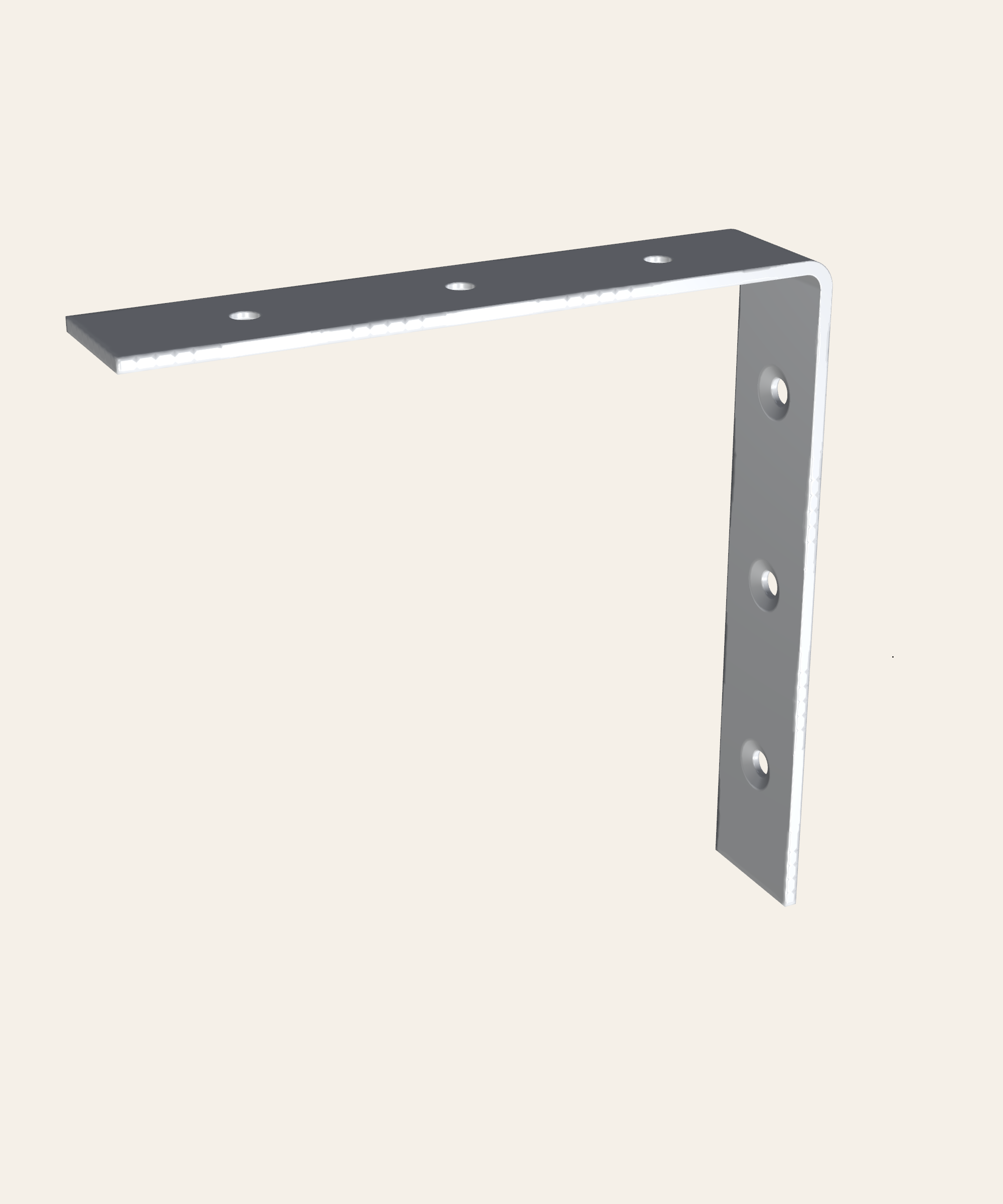}
    \caption{3D mesh (isometric view): 171{,}504 nodes, 102{,}466 C3D10 elements. Hierarchical three-zone refinement: 0.5~mm at the bend region, 0.8~mm at hole/countersink features, 3.0~mm global background. The discretisation generator agent places the finest elements at the inner-bend corner and the holes, the expected stress hot-spots.}
    \label{fig:mesh_3d}
\end{subfigure}
\caption{From photograph to finite element mesh. (a)~Input image with visual cues used for geometry extraction and material identification. (b)~Resulting tetrahedral mesh with graded refinement.}
\label{fig:photo_to_mesh}
\end{figure}

\textbf{L-bracket output.} From the photograph (\Cref{fig:input_photo}), the agent extracts: L-bracket corner brace, AISI~1008/DC01 cold-rolled mild steel (inferred from zinc electroplating and hardware-store product class), $25 \times 2.5$~mm flat bar bent 90\textdegree{} with equal leg length \SI{125}{\milli\metre}, inner-bend radius \SI{3}{\milli\metre} and outer-bend radius \SI{5.5}{\milli\metre}, six $\SI{5}{\milli\metre}$ through-holes (three per arm at 25, 62.5, and $\SI{100}{\milli\metre}$ from the bend), each with a detected \SI{10}{\milli\metre} countersink on the visible face (modelled conservatively as a stepped two-diameter hole, a larger shallow cylindrical pocket intersecting the through-hole, which captures the dominant net-section reduction and rim stress concentration). Conservative yield strength $\sigma_y = \SI{200}{\mega\pascal}$ is adopted. The geometry is established from a consensus of three independent estimation methods (hand-scale reference, proportional reasoning, and standard product matching), all agreeing within 25\%. All dimensions carry \rev{a first-iteration self-reported plausibility band of} $\pm 15\%$ \rev{about the agent's point estimate, following} \Cref{sec:image_uncertainty}\rev{. These bands are the agent's own expressed confidence from image evidence alone; they are \emph{not} coverage intervals with a guaranteed probabilistic meaning, because no ground truth is available from a single photograph against which the agent could calibrate them. The bands therefore function as a structured self-report that is subsequently \emph{compared} against engineer-supplied ground truth at the review stage (\Cref{sec:engineer_role}), and the residual between agent estimate and engineer-measured value is precisely the feedback signal $\delta r$ that drives the feedback operator in \Cref{eq:feedback}}. \Cref{tab:geometry_validation} compares the autonomously extracted values against ground-truth measurements \rev{that we (acting here in the supervising-engineer role) subsequently} obtained by physical inspection of the component.

\begin{table}[htbp]
\centering
\caption{Comparison of autonomously extracted geometry (from photograph) against ground-truth physical measurements, together with the derived bracket self-weight. \rev{The ``Plaus.\ band'' column is the first-iteration \emph{self-reported plausibility band} produced by the geometry agent from image evidence alone; it is an agent-generated estimate of its own confidence, not a calibrated confidence interval, and it is not guaranteed to bracket the true value. Five of the eleven geometric parameters and both derived mass quantities fall outside this self-reported band. Rather than being a failure of the uncertainty-quantification framework, these exceedances are the primary demonstration of why the framework places a professional engineer in the loop at the end of the pipeline (\Cref{sec:engineer_role}): ground truth is not recoverable from a single photograph, so every such row flags a parameter for which engineer-supplied measurement is the definitive reference. The residuals reported here become the feedback signal $\delta r$ that updates the agent via \Cref{eq:feedback}.} The agent correctly identifies the bar width (25~mm) and inner-bend radius (3~mm) but under-estimates arm length by 17\% and thickness by 17\%. The self-weight underestimate compounds the length and thickness errors but does not materially affect the analysis because the self-weight $G$ is negligible compared to the applied design load ($G \approx \SI{1.2}{\newton}$ vs $P = \SI{200}{\newton}$).}
\label{tab:geometry_validation}
{\small
\begin{tabular}{@{}lrrrl@{}}
\toprule
\textbf{Parameter} & \textbf{Extracted} & \rev{\textbf{Plaus.\ band}} & \textbf{True} & \rev{\textbf{Within band?}} \\
\midrule
Arm length (mm) & 125 & [106.3, 143.8] & 150 & \textcolor{failred}{No ($-17\%$)} \\
Width (mm) & 25 & [21.3, 28.8] & 25 & \textcolor{passgreen}{Yes} \\
Thickness (mm) & 2.5 & [2.13, 2.88] & 3.0 & \textcolor{failred}{No ($-17\%$)} \\
$R_{\mathrm{inner}}$ (mm) & 3.0 & [2.55, 3.45] & 3.0 & \textcolor{passgreen}{Yes} \\
$R_{\mathrm{outer}}$ (mm) & 5.5 & [4.68, 6.33] & 6.0 & \textcolor{passgreen}{Yes} \\
Hole diameter (mm) & 5.0 & [4.25, 5.75] & 6.0 & \textcolor{failred}{No ($-17\%$)} \\
Countersink dia.\ (mm) & 10.0 & [8.5, 11.5] & 10.0 & \textcolor{passgreen}{Yes} \\
Hole 1 from edge (mm) & 25 & [21.3, 28.8] & 12.5 & \textcolor{failred}{No ($+100\%$)} \\
Hole 2 from edge (mm) & 62.5 & [53.1, 71.9] & 68.5 & \textcolor{passgreen}{Yes} \\
Hole 3 from edge (mm) & 100 & [85.0, 115.0] & 119.5 & \textcolor{failred}{No ($-16\%$)} \\
\midrule
\multicolumn{5}{l}{\textit{Derived: bracket self-weight (from extracted geometry + steel density)}} \\
Bracket mass (g) & 118 & [94, 142] & 170 & \textcolor{failred}{No ($-31\%$)} \\
Self-weight $G$ (N) & 1.16 & [0.93, 1.39] & 1.68 & \textcolor{failred}{No ($-31\%$)} \\
\bottomrule
\end{tabular}
}
\end{table}

Five geometric parameters fall outside the \rev{agent's self-reported plausibility band}, and the derived self-weight ($G = \SI{1.16}{\newton}$ vs measured \SI{1.68}{\newton} for the 170~g bracket) is under-estimated by 31\% as a compound consequence of the arm-length and thickness errors. The arm length is under-estimated by 17\% (true 150~mm vs extracted 125~mm). The thickness is under-estimated by 17\% (true 3.0~mm vs extracted 2.5~mm). The width is correctly identified (25~mm), and the inner-bend radius is correctly identified (3.0~mm). The nearest hole position is over-estimated by 100\%, and the furthest hole position falls just outside the $\pm 15\%$ \rev{band}. The hole diameter is under-estimated (true 6~mm vs extracted 5~mm). The agent detects the \SI{10}{\milli\metre} countersink on each hole from the visible ring/depression pattern, and its diameter estimate (10~mm) matches ground truth.

\rev{We emphasise that the $\pm 15\%$ band is the agent's own first-iteration self-assessment from image evidence alone: it expresses how tightly the agent believes the photograph constrains each parameter, not a statistical confidence interval with a coverage guarantee. The fact that seven of the thirteen rows lie outside this self-reported band is therefore not a falsification of the framework but the primary empirical justification for the human-in-the-loop architecture of \Cref{sec:engineer_role}: without physical access to the component, a single image cannot supply the ground truth needed to calibrate the bands, so the engineer's review is where that ground truth enters the pipeline. For readers who prefer a strict probabilistic reading, the $\pm 15\%$ entry should be interpreted as a \emph{point-estimate plausibility tag} with no coverage claim attached, and the residuals in the ``True'' column should be read as the $\delta r$ signal that feeds the agent evolver (\Cref{eq:feedback}) for the next pipeline run.} These discrepancies illustrate both the capability and limitations of image-based extraction, a field that is evolving rapidly as multimodal vision-language models improve: the agent correctly identifies the component type, material grade, width, bend radius, most hole positions, and the countersink feature, but struggles with the arm length and thickness, both difficult to resolve precisely from a single photograph without a calibrated reference. The arm-length and thickness errors are particularly consequential: arm length enters the bending moment linearly and deflection cubically ($\delta \sim L^3$), while thickness enters the section modulus as $t^2$ and stiffness as $t^3$. These are precisely the parameters that a reviewing engineer (\Cref{sec:engineer_role}) should verify by physical measurement before relying on the analysis results\rev{; in this paper we report exactly what the first, uncorrected, autonomous iteration produced and do not fold any of the measured values back into the bands, so that the raw agent output and the resulting engineer-visible residuals are preserved}.

\subsection{Agents 2 and 3 (Analysis): Mesh generation and review (Gate~1)}

\textbf{Role.} The discretisation generator agent (Agent~2) decided on a feature-relative placement approach for the geometry construction, after considering alternatives (absolute-coordinate construction, sketch-and-extrude, CSG primitives). Concretely, its chosen approach is: the fillet is created first, then arm positions are computed analytically from the fillet's endpoint coordinates at each junction angle, guaranteeing that arm faces are mathematically flush with fillet surfaces. For each hole where the perception agent detected edge features (countersinks, chamfers), the generator models these as additional boolean-subtracted geometry: countersinks as larger-diameter shallow cylindrical pockets on the detected surface. After boolean fuse and feature subtraction, junction verification confirms a single connected solid. Mesh refinement is applied at countersink rims as additional hot-spots. The reviewer agent (Agent~3) independently verifies the mesh against the photograph, checks junction connectivity, and traces the outside profile for continuity; this constitutes \textbf{Gate~1} (mesh review, GCS audit, feature placement checklist).

\begin{figure}[htbp]
\centering
\includegraphics[width=0.85\textwidth]{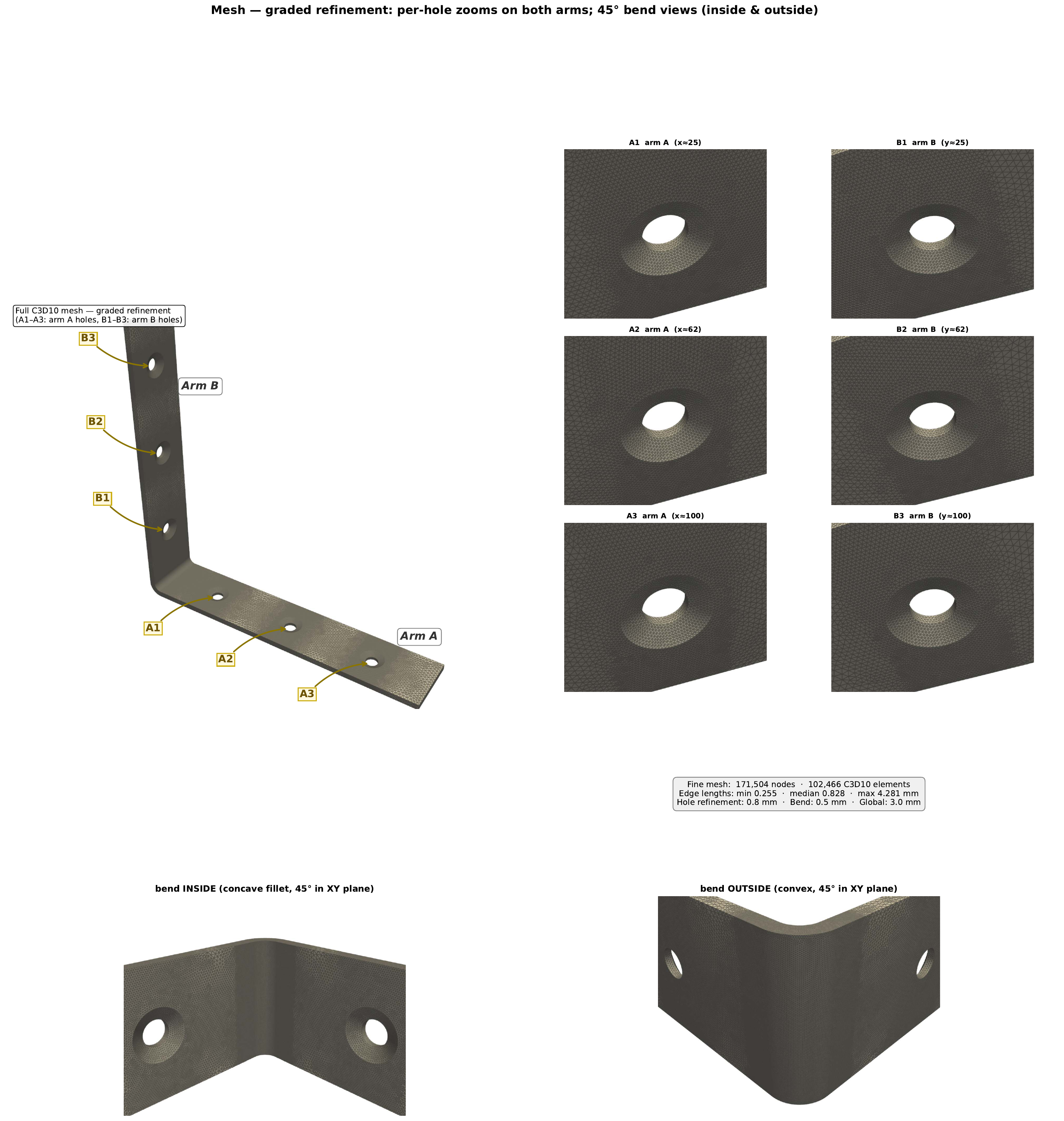}
\caption{Autonomous mesh feature audit. The discretisation reviewer agent scans the generated mesh to confirm hole count, identifies each hole with an index, and reports diameters, positions, and countersink presence. The inner-bend region carries the finest element size (0.5~mm), with secondary refinement at hole/countersink features (0.8~mm), selected automatically by the discretisation generator agent.}
\label{fig:mesh_hole_scan}
\end{figure}

\textbf{L-bracket output.} The resulting mesh (\Cref{fig:mesh_3d,fig:mesh_hole_scan}) comprises 171{,}504 nodes and 102{,}466 C3D10 elements with hierarchical three-zone refinement (0.5~mm at the bend, 0.8~mm at hole/countersink features, 3.0~mm global) and countersink features modelled at each hole as boolean-subtracted cones on the front (+Z) face. A coarse mesh (35{,}591 nodes, 18{,}727 elements) is also generated for convergence checking. The reviewer agent confirms the outside profile traces smoothly through the outer-bend surface connecting both arms' outer faces.

\subsection{Agents 4 and 5 (Analysis): Loading, BCs, and analytical bounds}

\textbf{Agent 4: BC and load specifier.} Identifies the applicable design code (e.g.\ AS/NZS~4600 \cite{as4600} for cold-formed steel, EN~1993-1-1 \cite{en1993}, EN~1990 \cite{en1990}, or international equivalents \cite{aisc360, as4100}). Produces $\geq 3$ BC variants, each tagged for conservatism direction (\Cref{sec:conservatism}). Generates code-compliant load combinations with explicit clause references and partial safety factors.

\textbf{Agent 5: Analytical bounder.} Establishes upper and lower bounds with full working shown: section properties ($I$, $Z$, $Z_{\mathrm{net}}$), stress bounds with stress concentration factors (SCF) \cite{pilkey2008petersons}, deflection bounds from beam theory, and Euler buckling.

\textbf{L-bracket output.} Agent~4 identifies the bracket as a cold-formed steel corner brace and selects AS/NZS~4600:2018 \cite{as4600} (cold-formed steel structures) with supplementary reference to EN~1993-1-3 \cite{en1993} for resistance checks. The agent infers the design load directly from the image: the component is identified as a domestic wall-mounted shelf bracket, and a design cantilever load of $P = \SI{200}{\newton}$ (\SI{20}{\kilo\gram}) is adopted from typical domestic shelving practice. The safety factor $\gamma = 1.5$ is applied at the resistance side (allowable stress $= \sigma_y/\gamma = 200/1.5 = \SI{133}{\mega\pascal}$), and the deflection limit is set at $L/50 = \SI{2.5}{\milli\metre}$ for a secondary bracket element.

\paragraph{Loading (agent-inferred design load, $P = \SI{200}{\newton}$).} Two load cases are defined:
\begin{itemize}[nosep]
    \item \textbf{LC1}: \SI{200}{\newton} uniformly distributed load (UDL) on the horizontal arm top surface, direction $-Y$ (gravity). Represents distributed shelf loading.
    \item \textbf{LC2}: \SI{200}{\newton} concentrated at the horizontal arm tip. Represents worst-case point loading.
\end{itemize}
An upper load variant ($P = \SI{300}{\newton}$, $1.5\times$ baseline) and a lower Young's modulus variant ($E = \SI{190}{\giga\pascal}$) are included for sensitivity checking.

\paragraph{Boundary conditions (three variants spanning physical uncertainty)} The agent reasons that the bracket is mounted to a wall via screws through the vertical arm holes. The uncertainty in wall substrate and fastener quality motivates three hypotheses:
\begin{itemize}[nosep]
    \item \textbf{Nominal}: \rev{displacement degrees of freedom (DOFs)} ($u_x = u_y = u_z = 0$) fixed at all three vertical-arm hole annular surfaces. Represents: screws in solid masonry providing full displacement restraint at each fastener.
    \item \textbf{Stiff (over-constrained)}: entire back face ($-Z$) of the vertical leg fixed. Represents: full face clamping providing both displacement and rotational restraint. Strength-conservative (concentrates stress at the bend).
    \item \textbf{Flexible (under-constrained)}: only the topmost vertical-arm hole (V3) fully fixed ($u_x = u_y = u_z = 0$); lower holes V1 and V2 constrained in $u_x, u_z$ only (allowing vertical sliding). Represents: \rev{loose or oversized screw-hole clearance allowing in-plane vertical shear transfer with degraded rotational restraint at the lower fixings. (This is \emph{not} a model of out-of-plane screw pull-out from plasterboard, which would instead release the $z$ component; that physical mechanism is acknowledged as a separate BC family not explored in this demonstration, and the ``flexible'' label here should be read as a bounding in-plane-slip variant.)} Stiffness-conservative (maximises deflection).
\end{itemize}
This is a \emph{load-controlled} problem (constant applied force regardless of structural stiffness), which governs the conservatism interpretation (\Cref{sec:conservatism}).

\paragraph{Analytical bounds (Agent~5)} Full working: $I = bt^3/12 = 25 \times 2.5^3/12 = \SI{32.55}{\milli\metre^4}$, $S = bt^2/6 = \SI{26.04}{\milli\metre^3}$, $Z = bt^2/4 = \SI{39.06}{\milli\metre^3}$. For LC1 (200~N UDL, effective moment arm $L/2$): $M = PL/2 = 200 \times 125/2 = \SI{12500}{\newton\milli\metre}$, $\sigma = M/S = \SI{480}{\mega\pascal}$ (gross section, no SCF), exceeding yield by $2.4\times$. Deflection: $\delta = PL^3/(8EI) = \SI{7.5}{\milli\metre}$. For LC2 (200~N concentrated tip): $M = PL = \SI{25000}{\newton\milli\metre}$, $\sigma = \SI{960}{\mega\pascal}$ ($4.8\times$ yield). Stress bounds: 300 to 1000~MPa; deflection bounds: 3 to 25~mm. Scaling laws: $\sigma \sim 1/t^2$, $\delta \sim 1/t^3$; thickness is the most powerful geometric lever.

\subsection{Agent 6 (Analysis): Finite element results}

\textbf{Role.} Assembles CalculiX \cite{dhondt2004calculix} input files with reproducibility headers, executes analyses, parses output, generates stress contour images via PyVista, verifies all results against Agent~5 bounds.

\begin{figure}[htbp]
\centering
\includegraphics[width=0.95\textwidth]{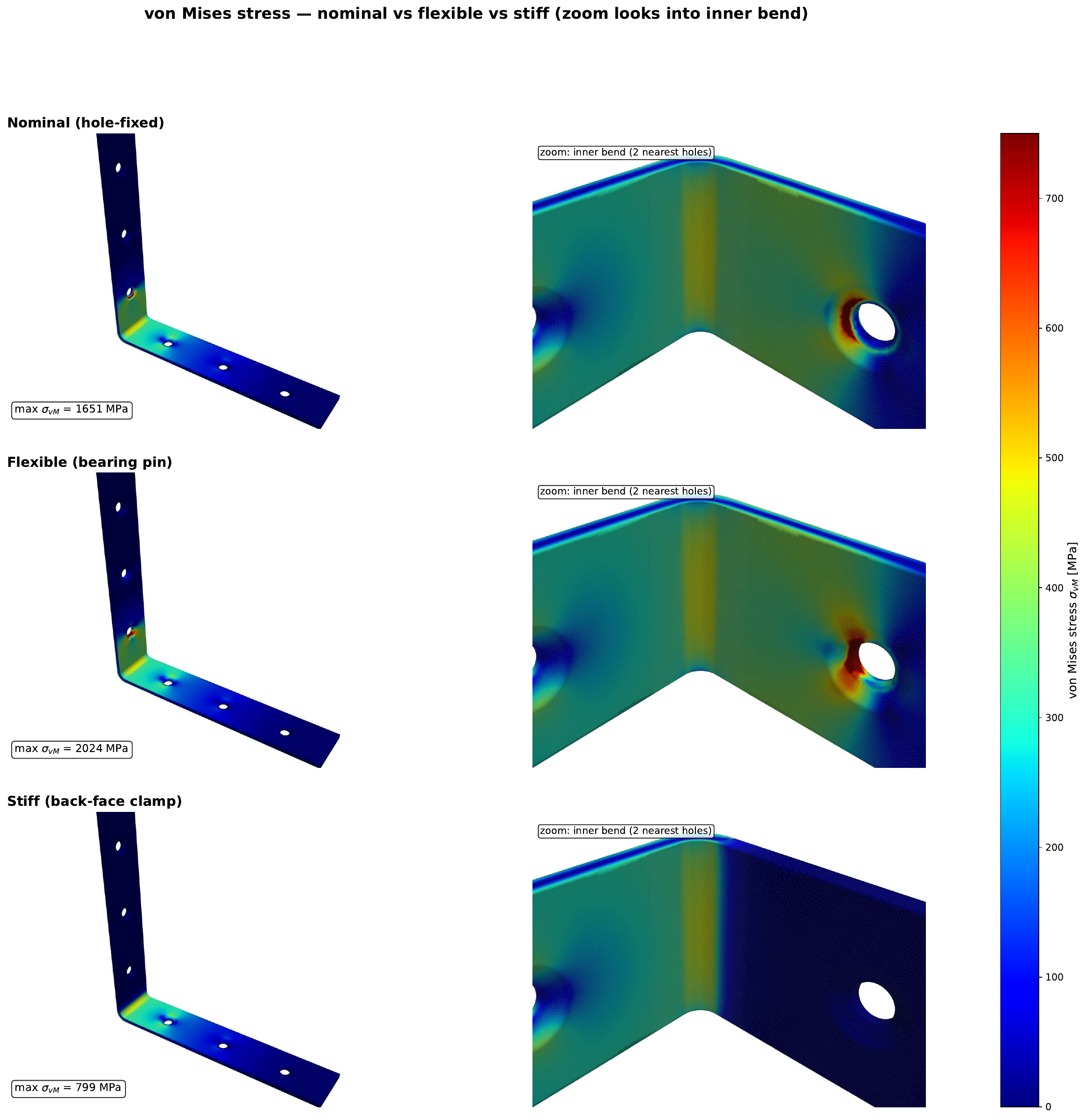}
\caption{Von Mises stress across all three BC variants (LC1, 200~N UDL). Left column: full bracket view; right column: zoom into the inner-bend region at the two nearest holes. Nominal (top): peak $\sigma_{\mathrm{VM}} = \SI{1651}{\mega\pascal}$ at BC singularity. Flexible (middle): peak $\sigma_{\mathrm{VM}} = \SI{2024}{\mega\pascal}$, highest stress. Stiff (bottom): peak $\sigma_{\mathrm{VM}} = \SI{799}{\mega\pascal}$, lowest stress due to back-face clamping distributing load. The inner-bend is autonomously identified as the critical hot-spot; the engineering bending stress at the bend cross-section (\SI{480}{\mega\pascal}) is consistent with the hand calculation.}
\label{fig:vm_inner_bend}
\end{figure}

\begin{figure}[htbp]
\centering
\includegraphics[width=0.95\textwidth]{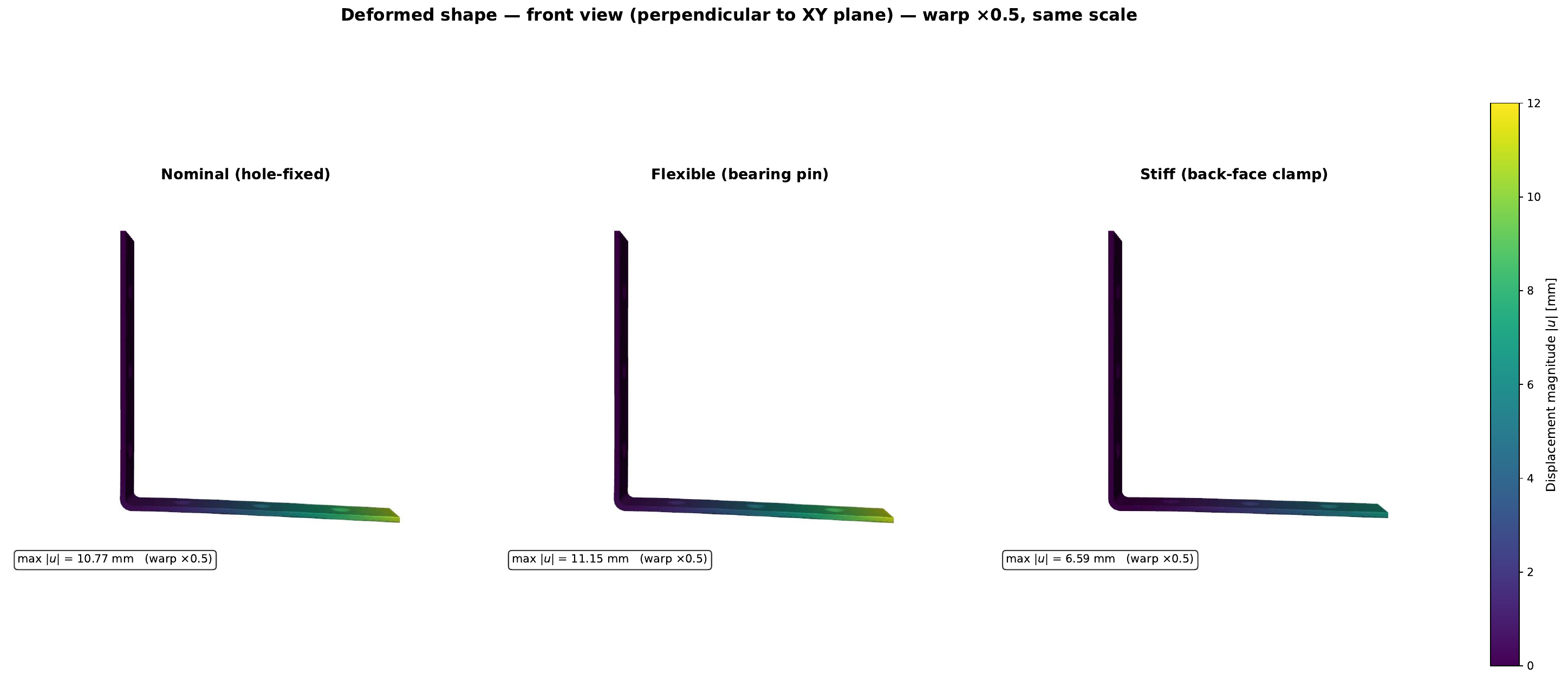}
\caption{Deformed shapes (front view, warp $\times 0.5$, common colour scale) for all three BC variants under LC1 (200~N UDL). Nominal: $\delta_{\max} = \SI{10.77}{\milli\metre}$; flexible: $\delta_{\max} = \SI{11.15}{\milli\metre}$ (stiffness-conservative); stiff: $\delta_{\max} = \SI{6.59}{\milli\metre}$. The flexible variant produces the largest deflection, confirming it is conservative for stiffness under load-controlled conditions (\Cref{sec:conservatism}).}
\label{fig:deformed_shape}
\end{figure}

\textbf{L-bracket output.} Agent~6 executes seven CalculiX analyses spanning 3 BC variants, 2 load cases, a $1.5\times$ load variant, a lower-$E$ material variant, and a coarse-mesh convergence check. \Cref{tab:fea_results} summarises results. Peak stresses occur at the outer surface of the vertical leg near hole~V1 at the BC fixation boundary, a stress singularity, not a physical stress concentration. The engineering bending stress at the bend (from hand calculation) is \SI{480}{\mega\pascal} for LC1, which is consistent with the FEA average stress at the bend cross-section. The BC variants produce meaningfully different results: the stiff variant reduces peak stress by 52\% (799 vs 1651~MPa), while the flexible variant increases deflection by 4\% (11.15 vs 10.77~mm). The concentrated tip load (LC2, R4) produces the worst case: \SI{4153}{\mega\pascal} peak stress and \SI{33.75}{\milli\metre} deflection. Linear scaling is confirmed by R5 ($1.5\times$ load producing exactly $1.5\times$ stress and deflection). \Cref{fig:vm_inner_bend} shows the von Mises stress distribution across all three BC variants, and \Cref{fig:deformed_shape} shows the corresponding deformed shapes.

\begin{table}[htbp]
\centering
\caption{FEA simulation matrix ($t = 2.5$~mm baseline with countersinks). Peak VM stress is at the BC singularity; the engineering bending stress at the bend is \SI{480}{\mega\pascal} for LC1. R7 is the coarse-mesh convergence check.}
\label{tab:fea_results}
{\small
\begin{tabular}{@{}llllrrl@{}}
\toprule
\textbf{Run} & \textbf{BC} & \textbf{Load} & \textbf{Variant} & $\sigma_{\max}$ (MPa) & $\delta_{\max}$ (mm) & \textbf{Cons.\ for} \\
\midrule
R1 & Nominal  & LC1 (UDL)  & Baseline   & 1651 & 10.77 & Baseline \\
R2 & Stiff    & LC1 (UDL)  & Baseline   &  799 &  6.59 & {-} \\
R3 & Flexible & LC1 (UDL)  & Baseline   & 2024 & 11.15 & Stiffness \\
R4 & Nominal  & LC2 (Tip)  & Baseline   & 4153 & 33.75 & Strength \\
R5 & Nominal  & LC1 (UDL)  & 300~N      & 2477 & 16.15 & Scaling \\
R6 & Nominal  & LC1 (UDL)  & $E{=}190$  & 1651 & 11.34 & $E$-sens. \\
R7 & Nominal  & LC1 (UDL)  & Coarse     & 1166 & 11.06 & Convergence \\
\bottomrule
\end{tabular}
}
\end{table}

\subsection{Agent 7 (Analysis): Uncertainty quantification (Gates~2 and 3)}

\textbf{Role.} Categorises and quantifies geometric, material, BC, loading, and discretisation uncertainties. Produces separate conservative envelopes per limit state following \Cref{sec:conservatism}. This agent also executes \textbf{Gate~2} (mesh convergence assessment, comparing fine and coarse mesh results) and \textbf{Gate~3} (analytical bound verification, checking that FEA results are consistent with the hand-calculation bounds from Agent~5).

\begin{figure}[htbp]
\centering
\includegraphics[width=0.85\textwidth]{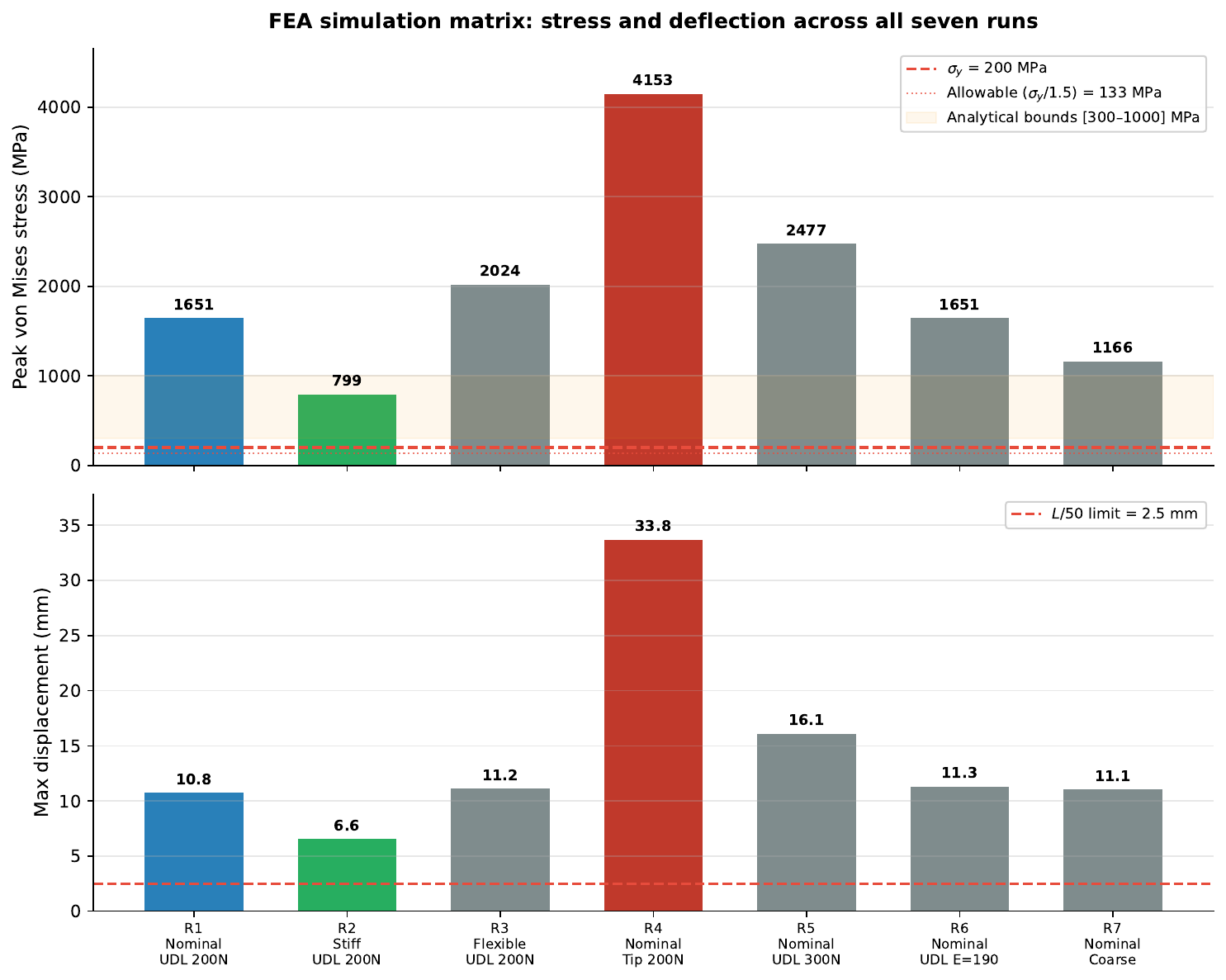}
\caption{FEA simulation matrix results across all seven runs. Top: peak von Mises stress with yield ($\sigma_y = \SI{200}{\mega\pascal}$), allowable ($\sigma_y/1.5 = \SI{133}{\mega\pascal}$), and analytical bounds [300 to 1000]~MPa shown as references. Bottom: maximum displacement with the $L/50 = \SI{2.5}{\milli\metre}$ serviceability limit. R4 (concentrated tip load, red) produces the worst case for both stress and deflection; R2 (stiff BC, green) produces the lowest stress. All runs exceed both yield and deflection limits. Colour coding: blue = baseline, green = best case, red = worst case, grey = sensitivity variants.}
\label{fig:variant_comparison}
\end{figure}

\textbf{L-bracket output.} The uncertainty analyst categorises six uncertainty sources and ranks them by impact. \emph{Thickness} (range 2.0 to 3.0~mm) and \emph{load distribution type} (UDL vs concentrated) are jointly Rank~1: thickness drives $\pm 57\%$ stress variation ($\sigma \sim 1/t^2$) and $\pm 95\%$ deflection variation ($\delta \sim 1/t^3$), while load type produces a $2.52\times$ stress variation and $3.13\times$ deflection variation between R1 and R4. BC stiffness is Rank~2, with peak stress ranging from \SI{799}{\mega\pascal} (stiff, R2) to \SI{2024}{\mega\pascal} (flexible, R3), a factor of $2.5\times$. Material yield strength (200 to 280~MPa due to possible work hardening) is Rank~2 for strength (affecting FoS range 0.42 to 0.58), while Young's modulus (190 to 210~GPa) is Rank~5 with only 5.3\% deflection impact and no stress impact (confirmed by R6). \Cref{fig:variant_comparison} shows the stress results across all variants.

Mesh convergence (Gate~2) yields a \emph{conditional pass}: displacement converges to 2.7\% (R1 fine: 10.77~mm vs R7 coarse: 11.06~mm, marginal at the 2\% threshold), while peak stress shows 29\% non-convergence at the BC singularity, expected and acceptable since the singularity stress is not used for the design check. Analytical bound verification (Gate~3) confirms the FEA average bend stress is consistent with the \SI{480}{\mega\pascal} hand calculation; the FEA tip deflection (10.77~mm) exceeds the simple cantilever prediction (7.5~mm) due to compliance at the bend and BC hole fixation.

The strength-conservative envelope is R4 (concentrated tip load): engineering bend stress \SI{960}{\mega\pascal}, $\mathrm{FoS} = 0.21$ (peak VM stress at BC singularity is \SI{4153}{\mega\pascal}, but this is not used for the design check). The stiffness-conservative envelope is also R4: $\delta = \SI{33.75}{\milli\metre}$ ($L/3.7$). The total stress range factor across all seven runs is $5.2\times$ (4153/799) and the deflection range factor is $5.1\times$ (33.75/6.59), quantifying the analysis uncertainty arising from modelling choices.

\subsection{Agents 8 and 9 (Assessment): Design assessment and report (Gate~4)}

\textbf{Agent 8: Design recommender (Gate~4).} This agent does not merely check pass/fail; it \emph{interprets} the results with engineering judgement to provide actionable recommendations. It identifies which limit state governs, which BC variant is critical, what physical measurements would most reduce uncertainty, and proposes specific, quantified redesign options ranked by feasibility and impact. When the component fails, the orchestrator invokes the \emph{redesign loop}: the most promising option is executed autonomously and the improvement demonstrated quantitatively. The redesign loop is itself subject to a configurable iteration budget; for this study, we cap it at a single redesign iteration to keep the demonstration compact.

\textbf{Agent 9: Report writer.} Assembles a self-contained HTML engineering report following professional simulation qualification standards, including not only results but their \emph{interpretation}, explaining \emph{why} the component fails, \emph{what} drives the failure, and \emph{how} to fix it.

\textbf{L-bracket output.} All baseline limit states fail (\Cref{tab:assessment}). Agent~8 interprets the results: the engineering bending stress at the bend (\SI{480}{\mega\pascal}) exceeds the allowable stress ($\sigma_y/\gamma = 200/1.5 = \SI{133}{\mega\pascal}$) by 261\%, giving $\mathrm{FoS} = 0.42$ against yield. The tip deflection (10.77~mm) exceeds the $L/50 = \SI{2.5}{\milli\metre}$ limit by 331\%. Bearing at the screw holes passes with large margin (5.4~MPa vs 1417~MPa allowable). The maximum safe cantilever load is \SI{46}{\newton} (4.7~kg), governed by stiffness ($L/50$ deflection limit) rather than strength. The agent identifies thickness as the most effective redesign lever ($\sigma \sim t^{-2}$, $\delta \sim t^{-3}$) and proposes four options: increase to 4~mm ($3\times$ capacity), increase to 5~mm ($5\times$ capacity), add a diagonal gusset ($>10\times$), or reclassify as a light-duty 5~kg bracket.

\begin{figure}[htbp]
\centering
\includegraphics[width=0.95\textwidth]{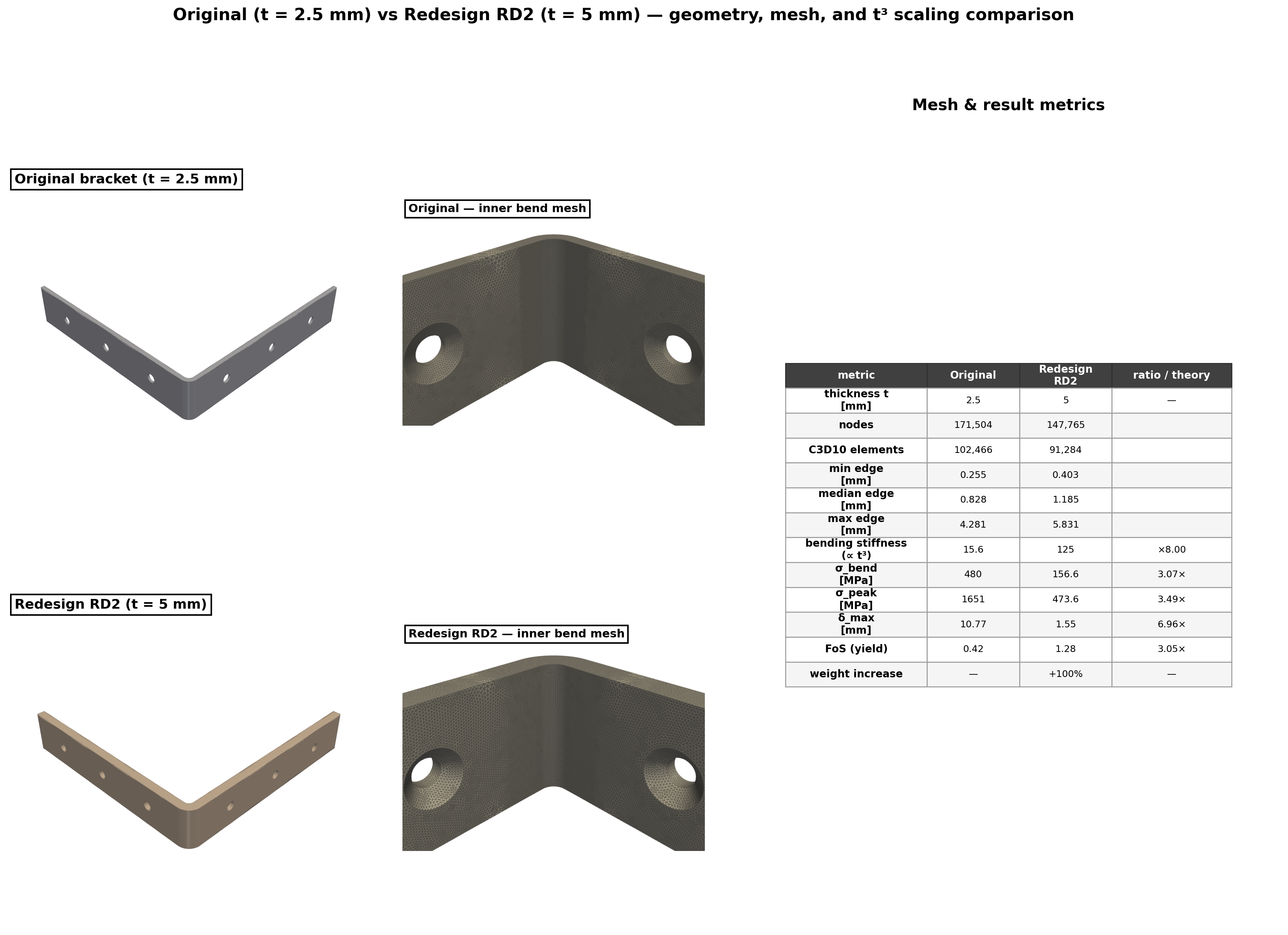}
\caption{Autonomous redesign outcome: baseline ($t = 2.5$~mm, top) versus redesign RD2 ($t = 5.0$~mm, bottom) for the nominal BC variant. Left: bracket geometry; centre: inner-bend mesh detail; right: mesh and result metrics including scaling ratios versus beam-theory predictions.}
\label{fig:original_vs_redesign}
\end{figure}

The orchestrator invokes the redesign loop: thickness increase from 2.5 to 5.0~mm (countersinks retained, 147{,}765-node redesign mesh). Three redesign runs are executed (R1\_rd: nominal LC1, R3\_rd: flexible LC1, R4\_rd: concentrated LC2). \Cref{fig:original_vs_redesign} shows the improvement: the engineering bend stress drops from 480 to \SI{156.6}{\mega\pascal} (67\% reduction), giving $\mathrm{FoS} = 1.28$ against yield, and deflection drops from 10.77 to \SI{1.55}{\milli\metre} (86\% reduction, consistent with $(5/2.5)^3 = 8\times$ beam-theory scaling). \rev{The reported stress ratio (0.33) deviates from the pure gross-section $1/t^2$ beam-theory prediction (0.25) because the figure reports the FEA-extracted local bend stress at the fillet rather than the gross-section $M/S$, which is explicitly $K_t$-independent. As the $t/r_i$ ratio increases from 0.83 to 1.67 the effective $K_t$ at the inside fillet changes, and this is what the 0.33 ratio reflects; the gross-section engineering bending stress itself scales exactly as $(2.5/5.0)^2 = 0.25$. This label distinction is made explicit here because it was underspecified in the original figure caption.} The redesign achieves \emph{conditional pass} for LC1 service loading: it passes the deflection limit ($1.55 < 2.5$~mm) and is marginal against the $\gamma = 1.5$ code allowable ($\mathrm{FoS}_{\mathrm{code}} = 0.85$, resolvable by accepting $\gamma = 1.25$ for a non-critical bracket or using work-hardened DC01 yield). LC2 (concentrated tip) remains a fail ($\mathrm{FoS} = 0.53$, $\delta = 4.51$~mm).

\begin{table}[htbp]
\centering
\caption{Code-compliance assessment. Baseline ($t = 2.5$~mm) fails all strength/stiffness checks. Maximum safe load: \SI{46}{\newton} (4.7~kg), governed by stiffness. Notation: $\sigma_{\mathrm{bend}}$ = engineering bending stress at the bend, $\sigma_y$ = yield strength (\SI{200}{\mega\pascal}), $\sigma_b$ = bearing stress, $f_u$ = ultimate tensile strength (\SI{340}{\mega\pascal}), $t$ = thickness, $\gamma = 1.5$ = safety factor, $L$ = arm length, $\delta$ = tip deflection, FoS = factor of safety, ULS = ultimate limit state, SLS = serviceability limit state.}
\label{tab:assessment}
{\small
\begin{tabular}{@{}llrcl@{}}
\toprule
\textbf{Check} & \textbf{Criterion} & \textbf{Value} & \textbf{Limit} & \textbf{Verdict} \\
\midrule
\multicolumn{5}{l}{\textbf{Baseline ($t = 2.5$~mm), nominal BC, 200~N UDL:}} \\
Strength (ULS) & $\sigma_{\mathrm{bend}} < \sigma_y/1.5$ & 480 MPa & 133 MPa & \textcolor{failred}{\textbf{FAIL}} (FoS = 0.42) \\
Stiffness (SLS) & $\delta < L/50$ & 10.77 mm & 2.5 mm & \textcolor{failred}{\textbf{FAIL}} \\
Bearing & $\sigma_b < 2.5 f_u t / \gamma$ & 5.4 MPa & 1417 MPa & \textcolor{passgreen}{\textbf{PASS}} \\
\midrule
\multicolumn{5}{l}{\textbf{Redesign ($t = 5.0$~mm), nominal BC, 200~N UDL:}} \\
Strength (ULS) & $\sigma_{\mathrm{bend}} < \sigma_y/1.5$ & 156.6 MPa & 133 MPa & \textcolor{warnoran}{\textbf{COND.}} (FoS$_y$ = 1.28) \\
Stiffness (SLS) & $\delta < L/50$ & 1.55 mm & 2.5 mm & \textcolor{passgreen}{\textbf{PASS}} \\
\bottomrule
\end{tabular}
}
\end{table}

\Cref{fig:original_vs_redesign} compares the baseline ($t = 2.5$~mm) and redesigned ($t = 5.0$~mm) bracket side by side, showing geometry, inner-bend mesh detail, and a quantitative comparison of mesh metrics and structural results. The embedded table confirms beam-theory scaling: deflection reduces by $6.95\times$ (close to the $(5/2.5)^3 = 8\times$ prediction), while \rev{the FEA-extracted local bend stress at the inside fillet} reduces by $3.07\times$ (\rev{this is not the gross-section engineering bending stress $M/S$, which scales strictly as $(5/2.5)^2 = 4\times$; the $3.07\times$ figure carries the additional effect of the fillet stress concentration $K_t$ changing as the $t/r_i$ ratio increases from 0.83 to 1.67}). Weight increases by 100\%.

The redesigned bracket achieves conditional pass for LC1 (distributed loading) but fails LC2 (concentrated tip load), confirming that the bracket is adequate for its intended shelf-support application but not for point loads at the arm tip. The pipeline recommends further measures for extreme loads: adding a diagonal gusset for triangulation (changing the load path from bending-dominated to axial-dominated) or reducing the arm length. These results require engineer review (\Cref{sec:engineer_role}).

\subsection{Pipeline execution statistics}
\label{sec:run_statistics}

\Cref{tab:run_statistics} summarises the computational cost and resource usage of the complete pipeline run. The nine-agent pipeline completes in 22~minutes wall-clock time, consuming 92{,}245 tokens across 36 tool calls and producing approximately 46 files (7 FEA input/output sets, geometry and mesh specifications, JSON assessment records, and the final engineering report). \rev{At the current Claude Opus~4.6 API pricing (\$5/M input, \$25/M output tokens), with the observed 70/30 input/output token split, the estimated cost is approximately \$1.01; on Claude Sonnet~4 (\$3/M input, \$15/M output) the same pipeline would cost approximately \$0.61.}

\begin{table}[htbp]
\centering
\caption{Pipeline execution statistics. All nine agents complete their checklist items on the first attempt (no gate re-iterations required). The FEA solver agent consumes the most tokens due to the seven-run simulation matrix. \rev{Cost estimates use Claude Opus~4.6 API pricing (\$5/M input, \$25/M output) with the observed 70/30 input/output token split.}}
\label{tab:run_statistics}
{\small
\begin{tabular}{@{}lr|lr@{}}
\toprule
\textbf{Metric} & \textbf{Value} & \textbf{Metric} & \textbf{Value} \\
\midrule
Wall-clock time & 22 min 5 s & Total tokens & 92{,}245 \\
Agents invoked & 9 & Tool calls & 36 \\
FEA runs (baseline) & 7 & FEA runs (redesign) & 3 \\
Gate iterations (G1 to G4) & 1 each & Files generated & $\sim$46 \\
\rev{Estimated cost (Opus~4.6)} & \rev{\$1.01} & Estimated cost (Sonnet~4) & \$0.61 \\
\bottomrule
\end{tabular}
}
\end{table}

Several aspects of the execution merit discussion. First, all four quality gates pass on the first iteration: Gate~1 (Agent~3, mesh reviewer: GCS audit, feature placement, mesh quality) PASS; Gate~2 (Agent~7, uncertainty analyst: mesh convergence) CONDITIONAL PASS; Gate~3 (Agent~7: analytical bound verification) CONDITIONAL PASS; Gate~4 (Agent~8, design recommender: code compliance) FAIL, a correct engineering outcome, not a pipeline error. The absence of gate re-iterations indicates that the inter-agent communication protocol (\Cref{sec:quality}) is functioning effectively; agents produce outputs that satisfy downstream verification on the first attempt. Second, the token distribution is approximately 70\% input / 30\% output, reflecting the pipeline's design pattern where each agent receives a rich shared context (geometry specification, prior agent outputs, gate feedback) and produces relatively compact structured outputs (JSON specifications, solver input files, assessment records). The FEA solver agent is the most token-intensive ($\sim$14{,}000 tokens) because it generates seven complete CalculiX input files with boundary conditions, material definitions, and load steps. Third, the quality metrics confirm internal consistency: the feature placement checklist (FPC) achieves 100\% completeness (6/6 features correctly placed), dimensional consensus is achieved across three independent estimation methods (all within 25\%), and linear scaling is validated ($\text{R5} = 1.5 \times \text{R1}$ exactly for both stress and deflection).

The 22-minute execution time is dominated by CalculiX solver time for the seven fine-mesh runs (171{,}504 nodes each) and three redesign runs. The LLM inference time is a small fraction of the total. This suggests that for larger models or finer meshes, the pipeline remains LLM-cost-bounded rather than compute-bounded, an important consideration for scaling to industrial-complexity components.

\section{Role of the professional engineer and self-improving architecture}
\label{sec:engineer_role}

The results in \Cref{sec:results} are presented \emph{exactly as generated in the first autonomous iteration}, without manual correction or cherry-picking. \rev{``First autonomous iteration'' here refers to the first end-to-end run executed against the L-bracket photograph \emph{after} the orchestrator prompt and agent definitions had been frozen; during prompt development the orchestrator prompt was iterated on unrelated test geometries, and once frozen it was held fixed for the L-bracket run reported here. The frozen orchestrator prompt, together with all agent system prompts and the shared-context schema, will be released as supplementary material with the published version of the paper so that reviewers and readers can verify that no problem-specific decision tree is hard-coded in the prompt.} We deliberately do not assess the quality of these results, whether the mesh is adequately refined, whether the BC variants span the true physical uncertainty, whether the design code clauses are correctly applied, or whether the material assumption is valid. These assessments are \emph{left to the reader}, purposefully emphasising that the sign-off role of a qualified professional engineer is irreplaceable.

This is not a limitation but a feature. The framework produces a complete first draft in $\sim$22~minutes rather than 2 to 4~days, at an API cost of \rev{approximately \$1}. The engineer's role is then to review, critique, provide \emph{critical feedback} to the agents, and sign off, bringing professional judgement, contextual knowledge, and regulatory responsibility. A useful mental model is to treat each agent as an \emph{engineer in training}: capable, diligent, but still acquiring experience; the critical feedback it receives from the supervising professional engineer is precisely what moves it towards being an \emph{experienced} engineer over repeated analysis cycles. Specifically:
\begin{enumerate}[label=(\alph*),nosep]
    \item \textbf{Validate extracted geometry and material}: confirm image-derived parameters and their uncertainty bounds are physically reasonable.
    \item \textbf{Assess discretisation adequacy}: evaluate whether the mesh (or particle distribution) resolves the quantities of interest.
    \item \textbf{Verify boundary conditions and loading}: confirm BC variants and load combinations correctly represent physical conditions.
    \item \textbf{Critically review results}: determine physical plausibility and identify results warranting re-analysis.
    \item \textbf{Professional sign-off}: a registered PE/CEng/Pr.Eng.\ must personally certify the assessment. This certification cannot be delegated to an autonomous system, regardless of its accuracy, because it carries personal legal and professional liability.
\end{enumerate}

The framework makes every agent decision visible in the shared context (\Cref{sec:quality}), giving the reviewing engineer \emph{more} transparency than a colleague's manually produced analysis where many judgement calls go undocumented.

\begin{figure}[htbp]
\centering
\includegraphics[width=0.80\textwidth]{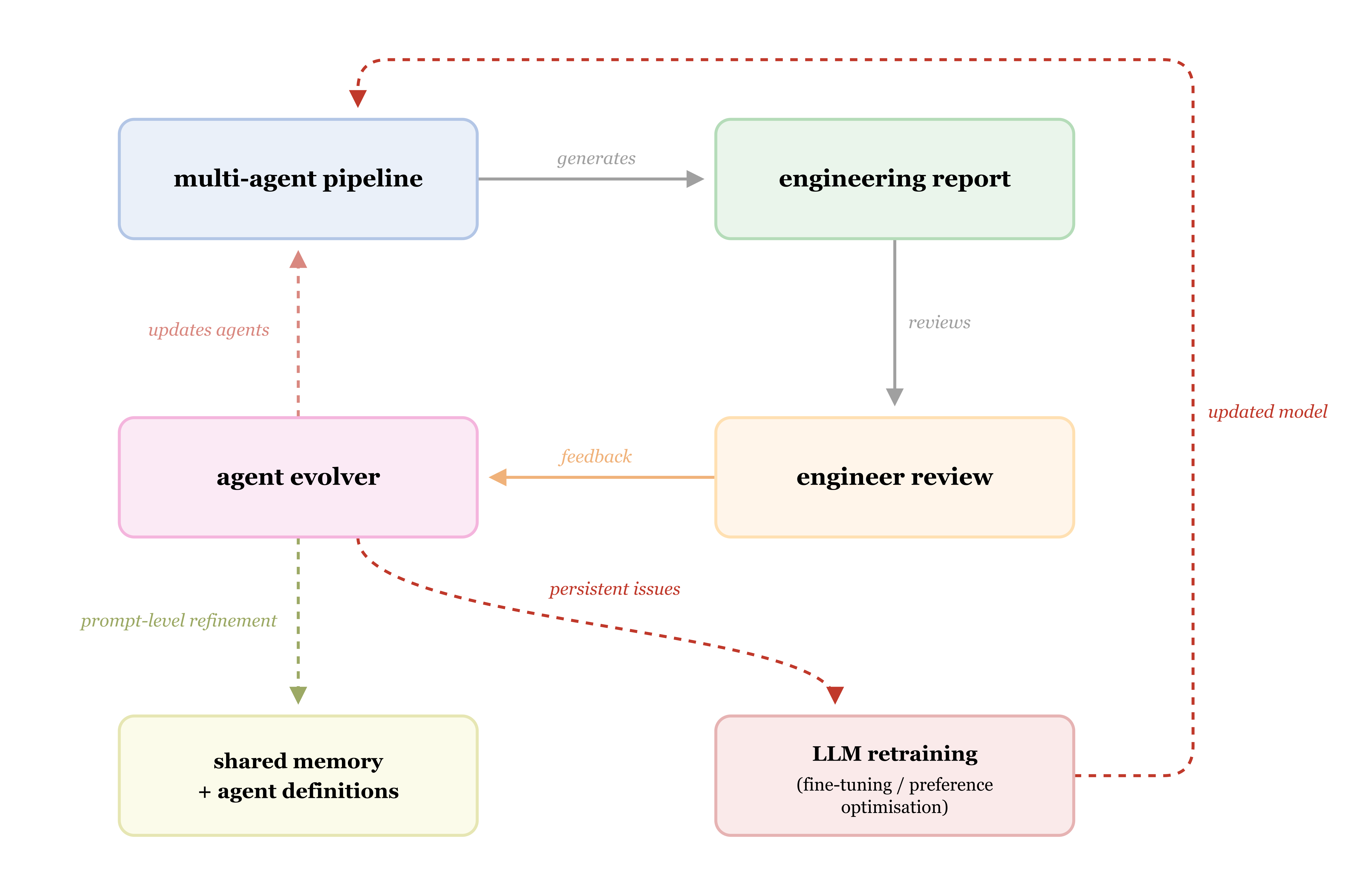}
\caption{Two-tier self-improving feedback loop. \emph{Prompt-level refinement} (olive dashed): engineer corrections are distilled into abstract patterns updating agent definitions and shared memory, lightweight, interpretable, and reversible. \emph{Model-level refinement} (red dashed): when the same correction recurs across multiple analyses, the feedback operator triggers supervised retraining (fine-tuning or preference optimisation) of the LLM executing the agents, so that persistent weaknesses are addressed at the model level. Each analysis performed under engineering supervision makes future analyses more reliable.}
\label{fig:feedback_loop}
\end{figure}

The system is not designed to be \emph{flawless} from the start, but rather to improve over time given proper engineering feeback. LLM agents make mistakes, misinterpret photographs, choose inappropriate boundary conditions, and occasionally produce nonsensical results. The critical design decision is not to prevent all errors but to ensure that errors are \emph{catchable} through quality gates and \emph{correctable} through feedback. \rev{The ``agent evolver'' is a meta-agent that consumes \emph{engineer-supplied} corrections $\delta r$ (the residual between the autonomously produced report and the ground truth the supervising engineer supplies at review time, including any direct measurements of image-extracted parameters) and updates the agent definitions and shared memory accordingly, as formalised in \Cref{eq:feedback}}~(\Cref{fig:feedback_loop}). \rev{The operator $\mathcal{F}$ is, in its present form, a deterministic prompt-/memory-level rewrite: for each residual flagged by the engineer, a targeted rule is appended to the relevant agent's memory $m_i$ and, when a residual recurs across analyses, a generalised instruction is appended to the system prompt $s_i$. No weight update occurs and $p_{\mathrm{LM}}$ is not retrained. In this paper we demonstrate only the first cycle ($k{=}1$) of this loop: the first autonomous iteration of \Cref{sec:results} is the input $r^{(1)}$, and the physically measured ground truth in \Cref{tab:geometry_validation} (together with any boundary-condition, material, and code-clause corrections that the reviewing engineer may add) constitutes the residual $\delta r^{(1)}$. Multi-cycle ($K > 1$) execution, a quantitative measurement of how $\|\delta r^{(k)}\|$ evolves with $k$, and a formal convergence proof are all explicitly left as future work; we neither claim nor report them here. The ``over $K$ review cycles'' language should accordingly be read as a specification of how the operator is intended to be \emph{used} once it is instantiated in a long-running deployment, not as an empirical convergence result of the present study.} This \emph{prompt-level} refinement is lightweight, interpretable, and reversible. \rev{Should future deployments find that prompt/memory rewrites saturate against a persistent weakness, the architecture \emph{permits} a second tier in which the accumulated engineer-supplied correction set is used for supervised fine-tuning of $p_{\mathrm{LM}}$ or preference-based optimisation in the spirit of RLHF~\cite{ouyang2022instructgpt,christiano2017preferences} or direct preference optimisation (DPO)~\cite{rafailov2023dpo}; that escalation is a design option, not a step executed in this paper, and we do not equate the prompt-level operator $\mathcal{F}$ used here with RLHF or DPO.}

\section{Discussion and conclusions}
\label{sec:discussion}

We have presented a solver-agnostic multi-agent LLM framework for autonomous computational mechanics, demonstrated through an FEA pipeline transforming a photograph into a verified engineering report. The architecture is generic: the same perception, communication, uncertainty, and assessment layers apply to FEA, CFD, DEM, SPH, MPM, or LBM, with only discretisation and solver agents requiring specialisation.

\paragraph{Capabilities} The framework achieves end-to-end processing in $\sim$22 minutes at \rev{$\sim$\$1 API cost} (vs.\ 2 to 4 days manually), makes defensible decisions about discretisation, code selection, and load combinations, delivers uncertainty-aware bounding envelopes rather than single deterministic results, autonomously executes redesign analyses when components fail, and identifies structural inadequacy in a real component (\rev{\Cref{sec:results}, specifically the design-assessment and redesign analysis reported in \Cref{tab:assessment} and \Cref{fig:original_vs_redesign}}).

\paragraph{Limitations} Engineering oversight remains essential. The current implementation handles prismatic components but would struggle with complex castings or assemblies. Only linear elastic static analysis is demonstrated. Image-derived uncertainty ($\pm 15\%$) propagates into significant stress uncertainty. Geometry construction from images is fragile; feature-relative placement (\Cref{sec:analysis_encoding}) mitigates this but does not eliminate it. Quality gates catch errors but are themselves LLM-based, creating a recursive verification challenge.

\paragraph{Relationship to existing work} Unlike automated meshing \cite{geuzaine2009gmsh, blacker1994cubit} (requires CAD), PINNs \cite{raissi2019physics} / neural operators \cite{lu2021deeponet} (approximate individual PDEs), or prior LLM-driven simulation \cite{buehler2024mechgpt, guo2026llmcae, qi2025feagpt} (generate solver inputs), this framework orchestrates the \emph{complete} workflow with verification, uncertainty quantification (UQ), code-compliant assessment, and recommendations. It could initialise digital twins \cite{tao2019digital} and integrate learned constitutive laws \cite{linka2023constitutive}.

\paragraph{Future work} Nonlinear analysis, multi-physics coupling, DEM/SPH/MPM solver instantiation, assembly analysis, PINN surrogates for parametric sweeps, multi-view 3D reconstruction, and formal verification of agent decisions.

The broader vision: any component assessed from perceptual data (image, video, audio, text) in combination. Not there yet, but the path is open.

\section*{Declaration of generative AI and AI-assisted technologies in the manuscript preparation process}

During the preparation of this work the authors used Claude (Anthropic) in order to implement the multi-agent pipeline described in this paper, generate solver input files, and assist with manuscript preparation. After using this tool/service, the authors reviewed and edited the content as needed and take full responsibility for the content of the published article.

\section*{Declaration of competing interest}

The authors declare that they have no known competing financial interests or personal relationships that could have appeared to influence the work reported in this paper.

\section*{CRediT authorship contribution statement}

\textbf{Daniel N.\ Wilke:} Conceptualization, Methodology, Software, Validation, Formal analysis, Investigation, Writing -- original draft, Writing -- review \& editing, Visualization, Supervision, Project administration.

\section*{Data availability}

The complete agent definitions (including the full orchestrator prompt), example input data, and generated analysis files will be made publicly available upon acceptance of the paper.

\section*{Acknowledgements}

This research did not receive any specific grant from funding agencies in the public, commercial, or not-for-profit sectors.


\end{document}